\documentclass{aa}
\usepackage{epsfig}
\begin{document}

\thesaurus{(...) }
\title{Very high energy gamma-rays from Centaurus X-3:
indications and implications}
\author{A.~M.~Atoyan \inst{1}\thanks{on leave from Yerevan Physics
Institute, Armenia }
\and P.~M.~Chadwick \inst{2}
\and M.~K.~Daniel \inst{2}
\and K.~Lyons \inst{2}
\and T.~J.~L.~McComb \inst{2}
\and J.~M.~McKenny \inst{2}
\and S.~J.~Nolan \inst{2}
\and K.~J.~Orford \inst{2}
\and J.~L.~Osborne \inst{2}
\and S.~M.~Rayner \inst{2}
}

\offprints{atoyan@hep.physics.mcgill.ca,  M.K.Daniel@durham.ac.uk }

\institute{ McGill University, Physics Department, Montreal H3A 2T8, Canada
   \and Physics Department, Durham University, Durham DH1 3LE, UK.}
\date{Received ...  accepted...}
\titlerunning{Gamma-rays from Cen X-3}
\maketitle

\begin{abstract}
We present the results of a detailed timing analysis of observations
of Cen~X-3 taken by the University of Durham Mark-6 imaging atmospheric
Cherenkov telescope in 1997-1999. The presence of a TeV $\gamma$-ray
signal at the overall $\geq 4.5\sigma$ significance level in the `fully cut'
image selected data, as reported earlier, is confirmed. A search for
possible modulations of $\gamma$-rays with the pulsar spin period
$P_{0}\approx 4.8 \,\rm s$ was performed by the step-by-step application of
image parameter cuts of gradually increasing hardness. The data of each of
23 days of observations have not revealed any statistically
significant Rayleigh power peak, except for 1 day when a peak with a chance
probability $p=6.8\times 10^{-7}$, was found in `soft-cut' data sets.
This modulation, if real, is blue shifted by 6.6 msec ($>10^3$
kms$^{-1}$) from the nominal second harmonic of the X-ray pulsar. Taking
the large number of frequency trials into account, the estimated final
probability of such a peak by chance still remains $< 10^{-2}$. Bayesian
statistical analysis also indicates the presence of such modulations. We show
that the behaviour of the Rayleigh peak disappearing in the fully cut data
set is actually quite consistent with the hypothesis of a
$\gamma$-ray origin of that peak. No modulation of the VHE
$\gamma$-ray signal with the pulsar orbital phase is found.

In the second part of the paper we consider different theoretical models
that could self-consistently explain the existing data from
Cen~X-3 in high-energy (HE, $E\geq 100\,\rm MeV$) and very high energy (VHE,
$E\geq 100\,\rm GeV$) $\gamma$-rays. We propose on the basis of the
energetics required that all reasonable options for the $\gamma$-ray
production in Cen~X-3 must be connected to jets emerging from the
inner accretion disc around the neutron star. One of the
principal options is a large-scale source, with
$R_{\rm s} \sim 10^{13} - 10^{14} \,\rm cm$\,; this assumes
effective acceleration of electrons up to $\sim 10 \,\rm TeV$ by
shocks produced by interaction of these jets with
the dense atmosphere of the binary. It is shown that such a quasi-stationary
model could explain the bulk of the $\gamma$-radiation features observed
except for the $\gamma$-ray modulations with the pulsar spin. These
modulations, if genuine, would require an alternative source with  $R_{\rm s}
\ll 10^{11}\,\rm cm$. We consider two principal models, hadronic and
leptonic, for the formation of such a compact source in the jet. Both models
predict that the episodes of pulsed $\gamma$-ray emission may be rather rare,
with a typical duration not exceeding a few hours, and that
generally the frequency of pulsations should be significantly shifted from
the nominal frequency of the X-ray pulsar. The opportunities to
distinguish between different models by means of future $\gamma$-ray
observations of this X-ray binary are also discussed.


\end{abstract}

\section{Introduction}

Centaurus X-3 has been one of the prominent galactic sources of hard
radiation since its discovery as one of the first cosmic X-ray sources
(Chodil et al. \cite{Chodil67}), and was the first X-ray pulsar to be
discovered in a binary system (Giacconi et al \cite{G71},
Schreier et al. \cite{Schreier72}). All the basic parameters of this
archetypal high mass X-ray binary are well known. The pulsar has a
spin period $P_0 \approx 4.8\,\rm s$ and an orbital period $P_{\rm orb}
\approx 2.1 \,\rm d $, with a gradual shortening (i.e. `spinning-up') of both
periods in time, and with a deep eclipse of the X-ray source at the
orbital phases $-0.12 \leq \phi \leq 0.12$ (for reviews see
Joss \& Rappaport \cite{Joss84}, Nagase \cite{Nagase89}). The X-ray
luminosity of the pulsar is very large, reaching $L_{\rm X} \sim
10^{38} \,\rm erg s^{-1}$ in the `high' state (e.g. White et al.
\cite{White83}, Burderi et al. \cite{Burderi00}), which implies a massive
accretion of material onto the neutron star from the optical
companion (Pringle \& Rees \cite{Pringle72}, Lamb et al. \cite{Lamb73}). The
optical companion (V779 Cen) was discovered by Krzeminski
(\cite{Krz74}) and has been identified as an evolved O-type star
with surface temperature $T\geq 3\times 10^4 \,\rm K$ at a distance $\sim
8\,\rm kpc$ from the Sun, and a bolometric magnitude of $M_{\rm bol} \sim -9$
(Hutchings et al. \cite{Hut79}). The masses of the stars in this binary are
estimated as $M_{\rm n}\simeq 1.2 \, M_{\odot}$ and
$M_{\rm O-star} \simeq 20\,\rm M_{\odot}$. The value of $\rm P_{orb}$
suggests the separation between the centres of the stars to be
$a=1.3\times 10^{12}\, \rm cm$ and the radius of the massive star
filling its Roche lobe $R_{\rm O} = 8.6\times 10^{11}\,\rm cm$
(see e.g. Clark et al. \cite{Clark88}, Ash et al. \cite{Ash99}).

Cen X-3 is also known as one of those X-ray binaries from which
$\gamma$-ray signals have been reported. In the domain of high energy (HE)
$\gamma$-rays, conventionally $E\geq 100 \,\rm MeV$, the $\gamma$-ray flux
during 2 weeks of observations of Cen X-3 in October 1994 by EGRET was at the
level
$F(>100\,\rm MeV) =(9.2\pm 2.3)\times 10^{-7} \,\rm ph\, cm^{-2} s^{-1}$
(Vestrand et al. \cite{V97}). This excess flux was significant at a
$5\sigma$ level, which is generally considered as a reliable detection with
the EGRET instrument. The timing analysis has shown a significant modulation
of the signal with the pulsar spin, precisely at the contemporaneous
frequency of X-ray pulsations measured by BATSE, with a chance probability
estimate based on the Rayleigh test statistic of about
$1.6\times 10^{-3}$. No modulation with the orbital phase of the
pulsar was found: the $\gamma$-ray signal seems to be quite homogeneously
distributed throughout the entire orbit, with 68 of the 264 HE $\gamma$-rays
from Cen X-3 being detected in the pulsar eclipse orbital phase $| \phi |
\leq 0.12$. This suggests a production site of the $\gamma$-rays far away
from the pulsar. The $\gamma$-ray signal was not found in the data of other
observing periods of Cen X-3 by EGRET, which suggests a significant
variability of the HE $\gamma$-ray source on a time scale of several
months (see Vestrand et al. 1997 for details).

In the domain of very high energies (VHE), conventionally $E\geq 100\,\rm
GeV$, sporadic $\gamma$-ray signals from Cen X-3 had been earlier reported by
the University of Durham (Brazier et al. \cite{Bra90a}) and the Potchefstroom
groups (North et al. \cite{N90}). Note that the early detections of VHE
$\gamma$-rays from X-ray binaries in the 80's had been carried out
by non-imaging Cherenkov telescopes which had rather poor sensitivity.
The $\gamma$-ray signals were therefore extracted mostly on the
basis of the timing analyses, and are subject to
controversy concerning their reliability (see Weekes
\cite{Weekes92}). It is worth noting in this regard that Cen X-3 is
the first and until now the only X-ray binary to be detected as a
source of $E \geq 400\,\rm GeV$ $\gamma$-rays by a contemporary imaging
instrument, the Durham Mark-6 telescope, as we have reported earlier
(Chadwick et al. \cite{C98,C00}). The excess ($\gamma$-ray) signal has been
found in the `ON-source' data during each of the 3 years of observations from
1997 to 1999. The estimated mean $\gamma$-ray flux is about  $F(>400\,\rm
GeV) \simeq 2.8 \times 10^{-11} \,\rm cm^{-2} s^{-1}$, at the
significance level for the entire data set of $4.7\sigma$. No
significant modulations in the combined ON-source data with either
the 4.8\,s pulsar period or 2.1\,d orbital period of the binary were found,
but these data were analyzed after application of the image cut procedures
(Chadwick et al. \cite{C00}) which we shall see later can be very
counter productive in a periodicity search.

In this paper we present the results of a more detailed timing
analysis of observations of Cen X-3, which in particular includes a
step-by-step search for possible short-term episodes of $\gamma$-ray emission
modulated with the spin of the pulsar, as well as a search for a modulation
of the $\gamma$-ray signal with the pulsar orbital phase (Section 2). Then we
carry out in Section 3 a detailed theoretical study of the consistency of
different models for the production of $\gamma$-rays in Cen X-3 with
the experimental data currently available in both HE and VHE domains.
Finally, in Section 4 we summarize the results of our current study,
and discuss some possible tests for future $\gamma$-ray observations
that could help to distinguish between different models of $\gamma$-ray
production in this X-ray binary.

\section{Timing analysis}

As we have reported earlier (Chadwick et al. \cite{C98,C00}), the entire set
of data of observations of Cen X-3 by the University of Durham Mark 6
telescope (see Armstrong et al. 1999 for a description) from
Narrabri, Australia, consist of pairs of 15 minute segments of
alternately ON- and OFF-source observations. A total of 129 ON-source
segments have passed the relevant consistency criteria in order to be
accepted for further analysis, yielding about $32 \,\rm hrs$ worth of data.
The data have been accumulated during 23 days of observations over 3 years
from 1997 to 1999. After application of a set of shower image
parameter cuts (see Chadwick et al. \cite{C00}), an excess of events
in the on-source data, indicating a possible presence of
$\gamma$-ray signal, is found in the data from each individual year.
In Table 1 we present the results of the analysis of the data for all of
the 23 days of observations. Note that the significance
$4.34\sigma$ in the overall data set for the image brightness parameter
$brightness > 800$ digital counts, which corresponds to the Monte-Carlo
estimated $\gamma$-ray energies $E> 400\,\rm GeV$, is somewhat lower than
$4.7\sigma$ as reported previously (Chadwick et al.\cite{C00}) because two
days worth of data were subsequently corrected for a problem with one of the
imaging PMT's.

For the periodicity analysis the air shower arrival times
were first corrected to the reference frame of the solar
system barycentre in order to exclude the timing effects connected
with the motion of the Earth.

\subsection{Orbital phase modulations}

Important information about a plausible site and mechanism
of gamma-ray production in X-ray binaries can be derived from
studies of possible correlations of the $\gamma$-ray fluxes with the orbital
phase of the pulsar. This can be particularly informative in the
case of Cen X-3 because it is an eclipsing binary, and because its optical
companion is a very luminous star. As shown by Bednarek (\cite{B00}), in the
case of $\gamma$-ray production near the orbit of Cen X-3, the optical depth
of the VHE $\gamma$-rays can vary from $\tau_{\gamma \gamma} \sim 10$ to
$\tau_{\gamma \gamma} < 1$ with the variation of the orbital phase from
$\phi \geq 0.12$  (i.e. outside the pulsar eclipse) to $\phi \sim 0.5$ due to
absorption on the optical/UV photons from the companion. If
$\gamma$-radiation is produced close to the pulsar then a strong correlation
of the signal strength with the orbital phase should be seen.

\begin{figure}[tbp]
\centerline{\epsfig{file=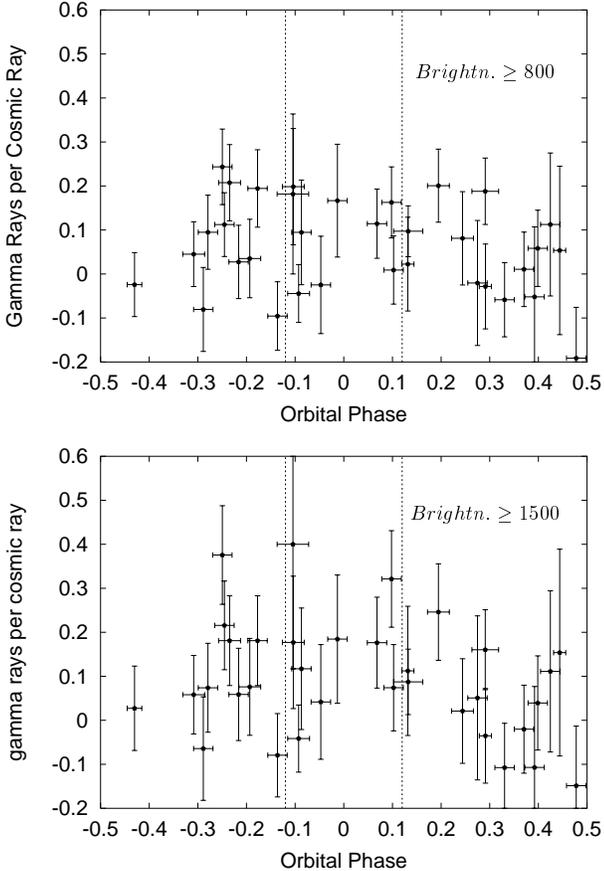, width=7.9cm}}
\caption{ON-OFF excess, in terms of `$\gamma$-rays per cosmic ray',
         in the fully cut data sets with 2 different brightness parameters,
         plotted with respect to orbital phases of observations in
         1997-1999.}
\label{Fig:Orbgrcr}
\end{figure}

For calculations of the orbital phase we use the ephemeris by Nagase et al.
(\cite{Nagase92}), which are in quite good agreement with the ones
by Kelley et al. (\cite{Kelley83}) we used earlier (Chadwick et al.
\cite{C98}, \cite{C00}), but being more contemporary are
more accurate and appropriate for the period of our observations. Because the
length of all ON/OFF segments accumulated during individual
observation nights varies and may reach several hours, i.e. a
significant fraction of the 2.1\,d orbital period, we have split the
data of the nights with a large ($\geq 5$) overall number of
on-source segments into sequences of 3-4 on-source segments (which in some
cases may overlap). In principle, such a grouping of data allows
one to also look for variations of the signal on timescales $\sim
1.5-2\,\rm h$, which could be of interest from a theoretical point of
view (see section 3 below).

A trend of a uniform distribution of the $\gamma$-ray signal with the
orbital phase of the pulsar is apparent in Fig.~\ref{Fig:Orbgrcr} where we
show the excess events in the on-source segments in terms of their ratio to
the number of events in the counterpart off-source segments. This method (see
Chadwick et al. \cite{C99}) allows an assessment of the strength of a
possible $\gamma$-ray signal with respect to the intensity of cosmic rays
independent of the daily variations in the performance of the telescope.
Note that there is some deficit in the number of on-source segments
corresponding to the orbital phase $-0.5 \leq \phi \leq -0.3$. For
all other phases, including the eclipse phase, the segments with
high (say $\geq 0.1$) ON-OFF excess are distributed quite homogeneously. It
is worth noting also that most of these high excess data points have
moved further up in significance when only the events with the
parameter $brightness \geq 1500$ digital counts are selected. A reasonable
explanation for this effect is that the $\gamma$-ray events are
indeed present in those segments, and that the spectrum of these
$\gamma$-rays is significantly harder than the spectrum of cosmic rays.

\subsection{Pulsar spin modulation}

In our previous analyses we were looking for a modulation of the
VHE $\gamma$-ray signal with the X-ray pulsar in a very narrow band
of frequencies around the fundamental frequency $\nu_{0}=1/P_0$
using the entire ON-source dataset. Meanwhile the earlier reports on the
$\gamma$-ray signals modulated with the X-ray pulsar spin in Cen X-3
(Brazier et al. \cite{Bra90a}), as well as in some other X-ray binaries
such as Cygnus X-3 (Brazier et al. \cite{Bra90b}) or Hercules X-1
(e.g. Dowthwaite et al. \cite{Dow84}, Lamb et al. \cite{Lamb88}), had
indicated that the pulsed $\gamma$-ray emission was sporadic, with a duration
typically not exceeding 1\,h and occasionally significantly shorter.
Moreover, the frequency of $\gamma$-pulsations reported in Her X-1
had on occasion been observed to be significantly shifted from the nominal
frequency of the X-ray pulsar by up to $\simeq 0.16\,\%$ (Lamb
et al. \cite{Lamb88}, Resvanis et al. \cite{Resvanis88}, Dingus et al.
\cite{Dingus88}).

Therefore it seemed worth conducting a search for any possible
modulation of the $\gamma$-ray emission with the BATSE derived
pulsar spin period in a wider bandpass around $P_0\approx 4.81\,\rm s$.
The test initially used was the standard Rayleigh test for
periodicity. The search was conducted at both the period and the half-period
(in the interval from 4.79 s to 4.83 s and 2.395 s to 2.415 s respectively)
due to the insensitivity of the Rayleigh test to light curves with a double
peak separated by $\pi$ in phase; a trait occasionally observed in Cen~X-3's
X-ray light curve (Tuohy \cite{T76}, Nagase et al. \cite{Nagase92}, Burderi
et al. \cite{Burderi00}). This feature suggests that both accretion hot spots
of the pulsar may become visible during one cycle of its rotation,
which is quite possible in the case of a relatively large inclination angle
of the pulsar magnetic axis. It is not improbable that the VHE
$\gamma$-rays might have the basic modulation near the half-period rather
than the full spin period of the pulsar.

The time range tested corresponds to Doppler shifted speeds of the putative
$\gamma$-ray source relative to the neutron star of up to $v\sim
1200 \,\rm km s^{-1}$. Note that the orbital speed of the Cen X-3 pulsar is
$414 \,\rm km s^{-1}$, and the speed of the wind driven by the optical
companion star in this binary is $\sim 1000\,\rm km s^{-1}$ (Clark et al.
\cite{Clark88}).

For a typical observation of Cen~X-3 during one night the duration
is $\Delta t_{\rm obs}\sim 3 \, \rm hr$ (in a succession of 'ON'
and `OFF' segments) therefore the chosen range of trial periods
corresponds to $\simeq 18-19$ independent Fourier frequency (IFF) intervals
$\Delta \nu = 1/\Delta t_{\rm obs}$ around $P_0$ (and about twice
that for the half-period search). Given the wide band of trial
periods studied, there is no need for an additional adjustment of
the event arrival times to the pulsar position because any
significant modulation frequency to be found could later on be
easily compared with the Doppler shifted frequency of the X-ray
pulsar at the known orbital position. Moreover, to adjust the event
times to the pulsar site may well be unjustified since both the HE and VHE
data give no indication of orbital modulation, with about a quarter
of all excess photons from the EGRET data being in the eclipse phase of the
pulsar and the episode of highest significance in the Mark~6 data occuring
during the X-ray eclipse. This implies the production site of the
$\gamma$-radiation to be far from the pulsar itself.

We have first considered, using the standard Rayleigh test
statistic, the data of each individual night of observation of Cen
X-3 after application of the full set of the Extensive Air Shower image
parameter cuts as described in Chadwick et al. (1998). This analysis did not
reveal any Rayleigh power that could be potentially significant
either around the main period $P_0\simeq 4.8\,\rm s$ or
around the second harmonic ($P\sim 2.4\,\rm s$). This, however, is
a result that one could reasonably expect.

The Rayleigh power at some frequency $\nu$ is defined as
$P_{\rm Rl} = (c^2 +s^2)/N$, where
$c=\sum_{j=1}^{N} \cos \theta_j$ and $s=\sum_{j=1}^{N} \sin \theta_j$,
and $\theta_{j}= 2\pi \nu t_{j}$ is the phase of the $j$-th event.
If the total number of events $N$ is large and the duration of the data set
is $\gg$ the test period $1/\nu$, then the sums $c$ and $s$ are
practically independent (only weakly correlated), so for simplicity
of further discussion we can take $P_{\rm Rl} \simeq 2 c^2/N$.
For uniformly  distributed events the mean $\overline{c}$
(and $\overline{s}$) is equal to 0. However the
root mean square (rms, the {\it dispersion}) of $c$ is not zero, but
is equal to $d=\sqrt{\overline{c^2}} =\sqrt{N/2}$ \footnote{It means
that $c$ and $s$ are expected to be in the range $\pm d$.},
so the expected mean $\overline{P_{\rm Rl}}=1$. If now there are
some $N_G$ events ({\it `$\gamma$-rays' }, G)
modulated at frequency $\nu$, and the rest $N_B=N-N_{G} $
represents the uniform background ({\it cosmic rays}, B) and $N$ is the total
number of events, then the expected mean Rayleigh power exceeds 1.
Taking into account that the
mean values $\overline{c} \sim \overline{s}\sim G/2$
(for sinusoidal modulations) and that $G\ll B\simeq N$,
one finds:
\begin{displaymath}
\overline{P_{\rm Rl}}\simeq 2 \frac{\overline{c_{B}^2}
+\overline{c_{B} c_{G}} +\overline{c_{G}^2} }{N_{G}+N_{B}}  =
1+\sigma_{0}^2\; ,
\end{displaymath}
where $\sigma_0 =N_G/\sqrt{2\, N_B} $. Note that, the latter variable
represents the statistical significance of the signal in the data,
which in practice is calculated as
$\sigma=\frac{(N_{\rm on}-N_{\rm off})}{\sqrt{N_{\rm on}+N_{\rm off}}}$,
which implies $N_{\rm on} = N_{B}+N_{G}$ and $N_{\rm off}=N_{B}$.

Any cuts made will affect the number of both $\gamma$-ray events,
$N_{G}^{\rm cut} =  F_{G} N_{G}$, and background cosmic ray events,
$N_{B}^{\rm cut}= F_{B} N_{B}$, where $F_G$ and $F_B$ are the respective
fractions of events surviving cuts. Since $F_G$ can be $\gg F_B$ imaging is a
very useful tool for background rejection. The significance of a signal,
however, is $\sigma \propto F_B^{-1/2}$; therefore the cuts made need to be
harsh to get the best signal to noise ratio, inevitably leading to a loss of
$\gamma$-ray events from the dataset. In a d.c. search this is acceptable
since merely increasing the observation time will lead to an increase in the
number of $\gamma$-rays in a dataset; however, in a periodicity search -
especially if the episodes of pulsed emission are short term and particularly
when combined with a low flux of VHE $\gamma$-rays  - any loss of
$\gamma$-ray signal could prove fatal to a positive detection. Whilst
cutting should improve the Rayleigh power mathematically, the low number
of events surviving cuts means the dispersion of the Rayleigh power about
the mean would give no reliable estimate of the true Rayleigh power, being
typically within the range $P_{{\rm Rl}} \simeq P_B + \sigma_0^2 \pm
2\sigma_0\sqrt{P_B}$; where the sign of the last term depends on whether
$c_B$ and $c_G$ are `coherent' (have the same or different signs). In
order to get a more robust statistic the cuts need to be relaxed, to
allow for the statistical fluctuations of $N_B$ and $N_G$.

As an example, taking the average `on-source' observation time to be 1.5
hours, and using even the mean computed flux of 1997,
$F(>400\,\rm GeV) = 5\times 10^{-11} \,\rm cm^{-2} s^{-1}$, which is a factor
of 2 higher than the mean during 1997-1999 from Cen~X-3, the mean number of
$\gamma$-rays for a detector with a collection area of $10^9$~cm$^2$ and 50\%
efficiency (Chadwick et al. \cite{C00}) is $\sim$130. Monte Carlo simulations
indicate that the fraction of $\gamma$-rays passing full image parameter cuts
is $\sim$20\%, so only 25-30 of these could be pulsed $\gamma$-rays. Note
that this is significantly less than minimum $40-50$ which is usually needed,
given the impact of statistical fluctuations, for a successful {\it
practical} application of the Rayleigh statistics, even if we neglect the
background events. Meanwhile, if one takes into account also that there are
on average about 400 events remaining in each night's data after the
application of imaging cuts, the {\it maximum} expected possible Rayleigh
power (assuming  $c_{G}^2+s_{G}^2 \simeq N_{G}^2$) would only be
$\overline{P_{\rm Rl}}^{\rm max} \simeq 1 + (25)^2/400 \sim 2.6$. This
would correspond to a chance probability of only  $exp(-P_{\rm Rl}) \sim
0.07$ - before the number of trials has been taken into account. Compare this
to the maximum Rayleigh power achieved by leaving 130 $\gamma$-ray events in
a background signal of $\sim$ 1000 events, found by relaxing the image
parameter cuts, of 17.9 that would give a chance probability of $\sim
10^{-8}$.

\begin{table*}[htbp]
\caption[]{The numbers of events in the `ON' and `OFF' pairs of
segments, the significance and the relative excess (in terms of `gamma-rays
per cosmic ray') of the signal in the data of 23 individual
nights of Cen X-3 observations after application of the full set of
the shower image parameter cuts, and for different brightnesses
of the images.}
\begin{center}
\renewcommand{\arraystretch}{1.5}
\begin{tabular}{|c|cccc|cccc|cccc|}   
\hline
 & \multicolumn{4}{c}{brightness $>800$ digital counts} \vline&
   \multicolumn{4}{c}{brightness $>1500$ digital counts} \vline&
   \multicolumn{4}{c}{brightness $>2000$ digital counts} \vline\\
\hline
Date & on & off & sig & gr/cr &on & off& sig & gr/cr & on & off & sig & gr/cr\\
\hline
01/03/97&446&401&1.546&0.112&265&218&2.139&0.216&199&173&1.348&0.150\\
03/03/97&572&515&1.729&0.111&391&329&2.311&0.188&339&282&2.287&0.202\\
04/03/97&383&319&2.416&0.201&233&187&2.245&0.246&184&149&1.918&0.235\\
01/06/97&265&256&0.394&0.035&184&171&0.690&0.076&171&156&0.830&0.096\\
02/06/97&455&383&2.487&0.188&304&262&1.765&0.160&260&219&1.873&0.187\\
04/06/97&200&185&0.764&0.081&146&143&0.176&0.021&131&138&-0.427&-0.051\\
05/06/97&323&295&1.126&0.095&218&203&0.731&0.074&182&169&0.694&0.077\\
07/06/97&394&377&0.612&0.045&273&258&0.651&0.058&231&215&0.758&0.074\\
\hline
1997 Total&3038&2731&4.042&0.112&2014&1771&3.950&0.137&1697&1501&3.466&0.131\\
\hline
27/03/98&689&628&1.681&0.097&411&378&1.175&0.087&296&280&0.667&0.057\\
29/03/98&661&589&2.036&0.122&424&347&2.773&0.222&309&243&2.809&0.272\\
30/03/98&364&373&-0.332&-0.024&226&220&0.284&0.027&174&163&0.599&0.067\\
17/04/98&182&178&0.211&0.022&109&98&0.765&0.112&78&78&0&0\\
19/04/98&339&336&0.115&0.009&232&216&0.756&0.074&183&167&0.855&0.096\\
26/04/98&59&56&0.280&0.054&45&39&0.655&0.154&39&32&0.831&0.219\\
27/04/98&473&441&1.058&0.073&371&334&1.394&0.111&313&270&1.781&0.159\\
28/04/98&272&293&-0.883&-0.072&206&220&-0.678&-0.064&172&181&-0.479&-0.050\\
29/04/98&151&126&1.502&0.198&113&96&1.176&0.177&87&85&0.152&0.024\\
\hline
1998 Total&3190&3020&2.157&0.056&2137&1948&2.957&0.097&1651&1499&2.708&0.101\\
\hline
13/02/99&431&451&-0.673&-0.044&323&337&-0.545&-0.042&263&266&-0.130&-0.011\\
15/02/99&78&66&1.00&0.182&42&30&1.414&0.400&30&16&2.064&0.875\\
16/02/99&923&902&0.492&0.023&614&602&0.344&0.020&492&495&-0.095&-0.006\\
17/02/99&922&889&0.775&0.037&643&611&0.904&0.052&536&479&1.789&0.119\\
20/02/99&206&212&-0.293&-0.028&164&170&-0.328&-0.035&144&146&-0.117&-0.014\\
21/02/99&355&294&2.39&0.207&248&210&1.778&0.181&210&172&1.944&0.221\\
\hline
1999 Total&2915&2814&1.334&0.036&2034&1960&1.171&0.0378&1675&1574&1.772&0.064\\
\hline
\hline
Grand Total&9143&8565&4.344&0.067&6185&5679&4.646&0.089&5023&4574&4.583&0.098\\
\hline
\end{tabular}
\end{center}
\label{tab:}        
\end{table*}

Thus, the hard image cut parameters which maximize the DC signal become
destructive for the search of $\gamma$-ray modulations with the pulsar spin
period if the data of individual observational nights are to be analyzed. For
this purpose the priority should be given to keeping in the data as
many $\gamma$-ray events as possible, but trying at the same time to
reasonably reduce the CR background events. In order to do so, we have
applied in succession a set of significantly softer image parameter cuts,
removing at the first step from the initial data only the events with the
orientation parameter $\alpha \geq 45^{\circ}$ (to compare with the $\alpha
\leq 30^\circ$ kept in the fully-cut sets). This procedure reduces
by a factor of $\sim 3$ the number of CR induced events, but does not
significantly affect the $\gamma$-ray events. Analyses of these data
did not reveal any statistically significant (i.e. corresponding to chance
probability $p$ below e.g. $10^{-3}$) Rayleigh peak either around the
first or the second harmonic of the X-ray pulsar.

The next set of data has been prepared applying a set of `soft' image
parameter cuts: by confining the position of the image within the camera
(parameter {\it distance}); and by discriminating the images using the {\it
width} and the {\it eccentricity} parameters. This procedure can be rather
efficient for suppressing CR background by a factor of $\sim 8-10$, although
it may also remove some $\gamma$-ray events. Lastly, we
have also searched for pulsar periodicity signs in the events after
discriminating the images by their {\it brightness}, choosing only
those with $>1500$ digital counts as compared to those with
$brightness > 800$ digital counts. This procedure effectively
increases the energy threshold by a factor $\sim 2$, therefore it is
efficient for the timing analysis only if the $\gamma$-ray spectrum is much
harder than the differential spectrum of the local CRs with the power-law
index $\alpha_{\sc cr}\simeq 2.7$. This idea stems from earlier theoretical
predictions that the spectra of episodic $\gamma$-rays expected in the VHE
domain from X-ray binaries may be anomalously hard due to very significant
absorption on the thermal optical/UV photons produced either by the compact
$\gamma$-ray source (Aharonian \& Atoyan \cite{AA91}, \cite{AA96}) or by the
optical companion star in the particular case of Cen X-3 (Bednarek
\cite{B00}).

\begin{figure}[tbp]
\centerline{\epsfig{file=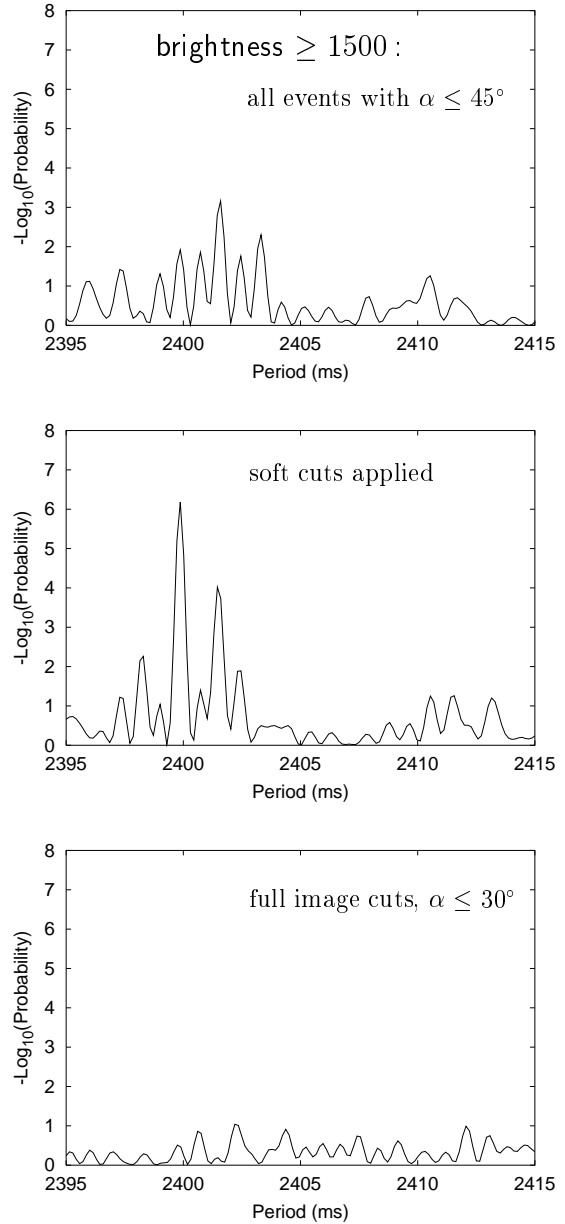, width=7.3cm}}
\caption{Rayleigh power probabilities in the search window around the
         half-period (the second harmonic) $P_0/2$ of the X-ray pulsar in the
         data sets of February 21, 1999 observations after application of
         image parameter cuts with different hardness.}
\label{Fig:RayHP1500}
\end{figure}

\begin{figure}[tbp]
\centerline{\epsfig{file=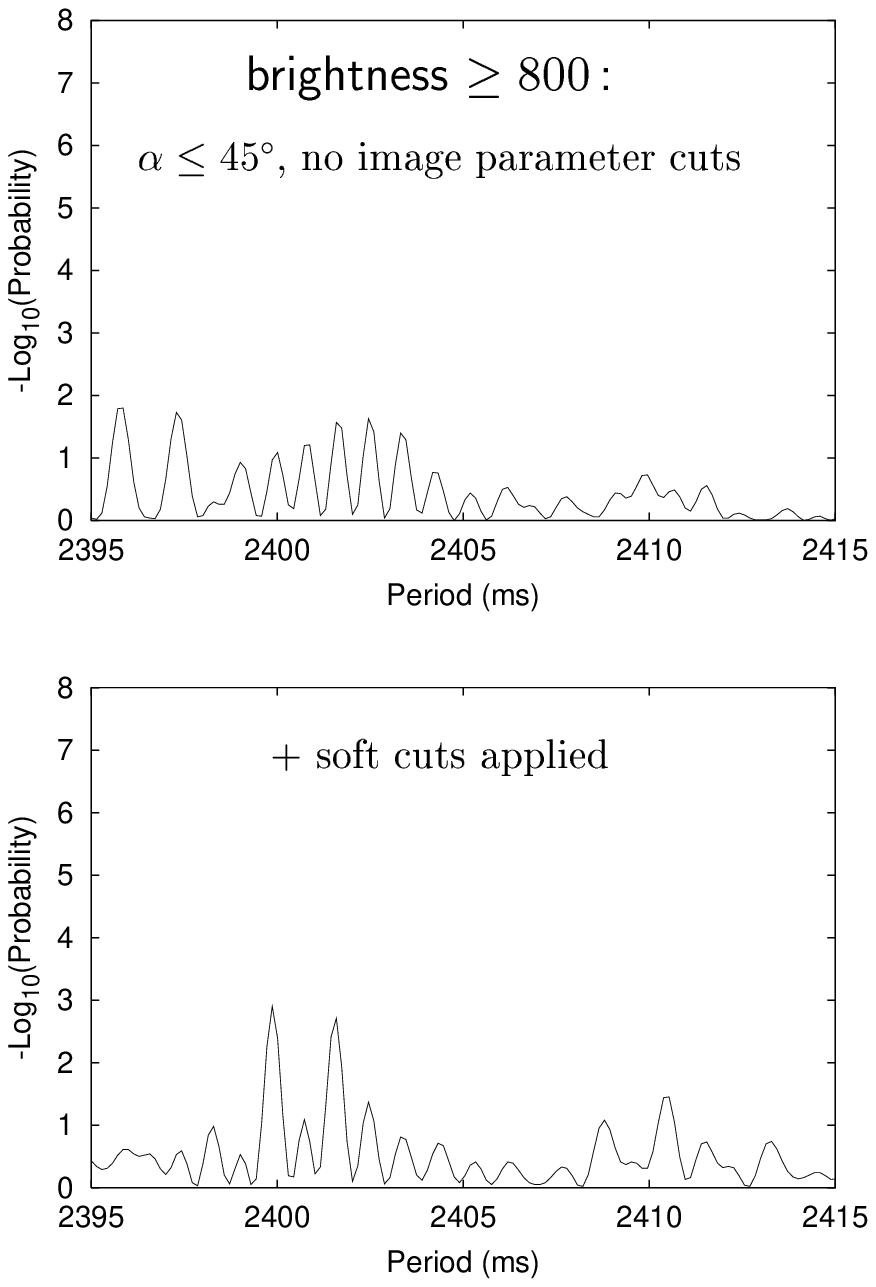, width=7.3cm}}
\caption{Same as in Fig.~\ref{Fig:RayHP1500}, but for the events with
         parameter $brightness > 800$ digital counts.}
\label{Fig:RayHP800}
\end{figure}

The detailed analysis of all these data sets has revealed a strong Rayleigh
power peak, corresponding to the chance probability for an artifact
modulation of only $p \sim 6.8\times 10^{-7}$ (see Fig.~\ref{Fig:RayHP1500}),
in the data of observations on 21 Feb, 1999 at $P_1=2.3999\,\rm s$. This
period is blue-shifted from the nominal second harmonic of the X-ray pulsar
($P_{0}/2 = 2.408785 \,\rm s$, after correcting to the pulsar orbital
phase $\phi \simeq 0.765 $ for that data) by $6.6 \,\rm msec$, and implies
a motion of the $\gamma$-ray source (if true) with respect to the neutron
star with a velocity $> 10^3 \,\rm km s^{-1}$. Interestingly enough, but not
surprisingly, this peak is found in the data after both the soft image cuts
and $brightness > 1500$ digital counts criteria have been applied. For
comparison, in Fig.~\ref{Fig:RayHP800} we show the results for the low energy
threshold case, $brightness > 800$ digital counts. The peaks
disappear in the fully cut data, which is not surprising in light of
the discussion above.

A simple estimate for an overall probability $p_{\rm tot}$ to find a
Rayleigh power $P_{\rm Ray}$ such that
$p_0=exp(-P_{\rm Ray})\leq 6.8\times 10^{-7}$ by chance can be derived
taking into account that for a mean duration of our observations of Cen X-3
per night is about 3\,hr, there are on average about 18.6
IFF intervals in the search window $4.79-4.83\,\rm s$ and twice that number
in the second harmonic search window, $2.395-2.415\,\rm s$. Because we have
prepared $6=3\times 2$ combinations of data (for different
parameter cuts and brightnesses) for each of the 23 days,  the
overall number of trials is $N_{\rm IFF}\leq 8\times 10^3$.
This means that the overall probability of this peak arising by
chance increases to $p_{\rm tot}\simeq N_{\rm IFF}\times p_0 \leq 5.4\times
10^{-3}$.

\begin{figure}[tbp]
\centerline{\epsfig{file=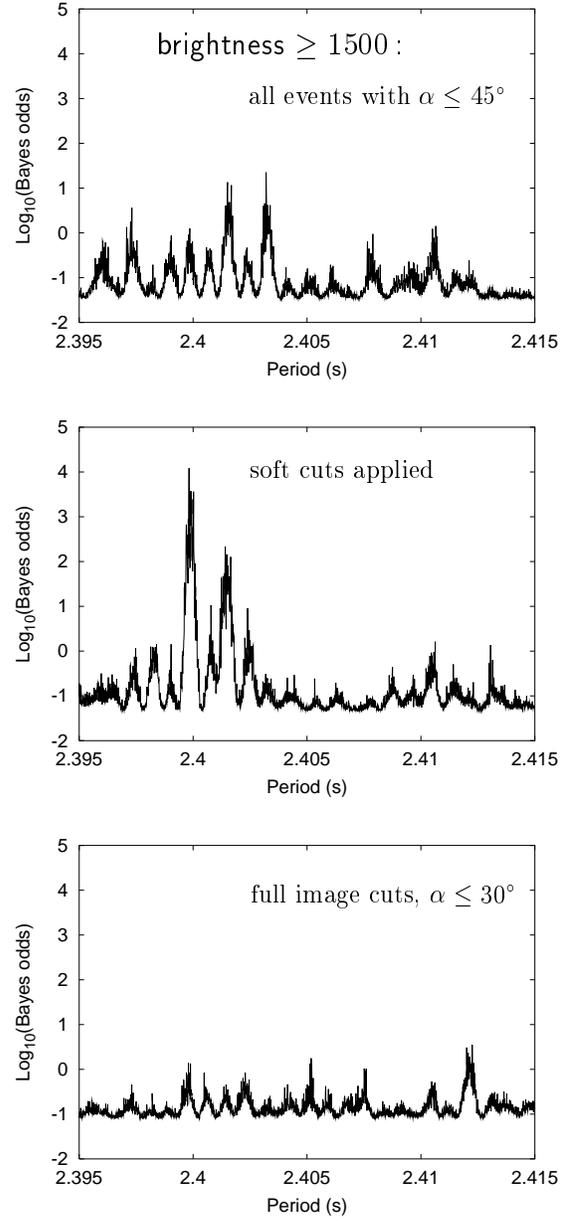, width=7.3cm}}
\caption{Similar to Fig.~\ref{Fig:RayHP1500}, but for the Bayesian
         odds analyses.}
\label{Fig:Bayes1500}
\end{figure}

Although the chance probability is still small it is not a conclusive
argument for VHE $\gamma$-ray periodicity. In an attempt to gain better
control of the hypothesis testing the analysis was repeated using a Bayesian
technique (see Gregory \& Loredo \cite{GL92}, Orford \cite{O00} for details).
This method applies Bayesian probability theory by comparing the phase
distribution of a constant model for the signal to members of a class of
models with periodic structure. The periodic models describe the signal plus
background with a stepwise function, resembling a histogram, of m phase bins
per period. In the case of a periodic model, the non-uniformity in phase is
characterised by the varying contents of the phase bins. Although the number
of phase bins needed to detect any light curve and the origin of phase are
unknowns, these 'nuisance' parameters can be marginalised (integrated out);
an important technical advantage of Bayesian inference over standard
frequentist statistics. Since an arbitrary postulated light curve may be of
any shape the method automatically applies Ockham's razor, in that models
with fewer variables are automatically favoured unless the evidence from the
data more than compensates for models of greater complexity. If there are
{\it m} phase bins the average rate $A=\frac{1}{m} \sum_{j=1}^m r_j$ and the
fraction of the total rate in phase bin {\it j} is $f_j=\frac{r_j}{mA}$. The
likelihood function is shown to reduce to
\begin{displaymath}
P(D|\omega,\phi,A,{\mathbf f}, M_m)= \Delta t^N (mA)^N e^{-AT}
\left(\prod_{j=1}^m f_{j}^{n_j}\right)
\end{displaymath}
where $\omega$ is the postulated angular frequency, $\phi$ the starting
phase, ${\mathbf f}$ the set of $m$ values of $f_j$ and $n_j$ being the
number of events occurring in bin $j$.

The joint prior density for the parameters $\omega, \phi, A, \mathbf{f}$ is
\begin{displaymath}
p(\omega, \phi, A, {\mathbf f}|M_m)=p(\omega|M_m)
                                    p(\phi|M_m)
                                    p(A|M_m)
                                    p({\mathbf f}|M_m).
\end{displaymath}
The prior densities are
\begin{itemize}
\item $p(\phi|M_m) = 1/2\pi$, this assumes any starting phase is equally
likely,
\item $p(A|M_m) = 1/A_{max}$, this assumes that $A$ does not change
during the observation and any value of $A$ from $A = 0$ to $A = A_{max}$
is possible,
\item $p(\omega|M_m) = \frac{1}{\omega ln(\omega _{hi}/\omega _{low})}$,
where $[\omega _{hi},\omega _{low}]$ is the least informative prior for
the range of $\omega$,
\item $p({\mathbf f}|M_m) = (m-1)!\delta (1-\sum_{j=1}^m f_j)$, where
$\delta$ denotes the dirac $\delta$-function.
\end{itemize}
The assignment of the priors beforehand is all that is needed before
comparing the likelihoods of the models. The two models are equally likely
{\it a priori} and so the prior likelihood of the non-periodic model ($M_1$)
is $p(M_1) = 1/2$ and that for a periodic model ($M_m, m=2, m_{max}$) is
$p(M_m|I) = 1/2\nu$, where $\nu = m_{max}-1$.

The final result for the odds $O$ in favour of a periodic model and against a
uniform model of phase when the phase and period are unknown (as in our case)
is
\begin{displaymath}
O_{m1} = \frac{1}{2\pi ln(\frac{\omega_{hi}}{\omega_{low}})}
         \frac{N!(m-1)!}{(N+m-1)!}
         \int_{\omega _{low}}^{\omega _{hi}}\frac{d\omega}{\omega}
         \int_0^{2\pi} d\phi \frac{m^N}{W_m(\omega,\phi)}
\end{displaymath}
where $W_m(\omega,\phi)$ is the number of ways that the set of $n_j$ counts
can be made by distributing $N$ counts in $m$ bins:
\begin{displaymath}
W_m(\omega,\phi) = \frac{N!}{\prod_{j=1}^m n_j!}
\end{displaymath}
and $n_j$, the number of events placed in the $j$th phase bin depends on
$\omega, \phi \;{\mathrm and}\; m$.
The overall likelihood of periodicity is then found from
\begin{displaymath}
O_{per} = \sum_{m=2}^{m_{max}} O_{m1} \; .
\end{displaymath}

The attractiveness of the Bayesian technique is that the overall probability
of the hypothesis of a time-modulated signal does not scale linearly with the
number of trials as in the Rayleigh test, but with the ln(period range
searched). The periodicity search is therefore only penalised by the fact
that the period is uncertain within a range and not by the number of times
that region is searched. In Fig.~\ref{Fig:Bayes1500} we show the Bayesian
odds results for the same data sets as used in Fig.~\ref{Fig:RayHP1500} for
the Rayleigh statistics. Formally, the peak in the median panel in
Fig.~\ref{Fig:Bayes1500} means that the hypothesis that the data set does
contain a signal modulated at the period $P_1 = 2.3998\,\rm s$ is $N_{\rm
odds} = 1.2\times 10^4$ times more probable than the hypothesis of a
uniform time distribution. Note that the positions and the qualitative
behaviour of the 2 highest peaks in Figures \ref{Fig:RayHP1500} and
\ref{Fig:Bayes1500} are similar.

Finally, it is worth noting the earlier reports in the 1980's about
the detection of anomalously pulsating episodic VHE $\gamma$-ray emission
from Her~X-1 by three independent groups (Whipple, Lamb et al. 1988;
Haleakala, Resvanis et al. 1988; and Los Alamos ,Dingus et al. 1988), offset
from the main period by $\sim$ 0.16\%. It is also worth noting that Her~X-1
had a short episode ($\Delta t \leq 1 \, \rm hr$) of $\gamma$-ray emission
reported as being detected simultaneously and independently by the University
of Durham (Chadwick et al. \cite{C87}) and the Whipple Observatory (Gorham et
al. \cite{Gor86}) groups, with both pulsation frequencies and estimated
fluxes being consistent in the initial uncut data of both groups. After
application of hard AZWIDTH parameter cuts which had earlier produced the
best results for the Crab Nebula signal (Weekes et al. 1989), the pulsations
in all Her X-1 data of the Whipple group disappeared (Reynolds et al.
\cite{Reynolds91}). This, however, is just the outcome that one would
expect, as we have discussed above in this Section. Taking into account
the rather steep spectrum of the Crab Nebula compared to the one
expected from Her X-1 (Aharonian \& Atoyan \cite{AA96}), we may here
hypothesise that an application of much softer parameter cuts, that would
first of all secure a larger number of possible $\gamma$-ray events in the
remaining data, could possibly have a better chance to enhance the initial
signal.

\section{Possible models}

\subsection{General considerations}

The results of observations of Cen X-3 in both HE and VHE $\gamma$-ray
domains allow us to significantly constrain different models of
$\gamma$-ray production in this X-ray binary. The first informative
result is the absence of any signs of orbital modulation of the $\gamma$-ray
signal in either of these energy domains. Because of the presence of a very
luminous O-type star close to the pulsar the VHE radiation produced at the
pulsar orbit would be significantly absorbed on the optical/UV photons from
the companion prior to escape from the system. Therefore the $\gamma$-ray
source cannot be placed close to the neutron star. This conclusion is also in
agreement with the fact of detection of $\gamma$-rays by EGRET (Vestrand et
al. \cite{V97}), and probably also by the Mark-6 telescope, in the phase of
complete eclipse of the X-ray pulsar.

Thus, at least for this particular binary any model assuming production
of $\gamma$-rays directly by the pulsar, or models invoking the accretion
disc around the neutron star as a site for $\gamma$-ray production
(e.g. Chanmugam \& Brecher \cite{Chan85}, Katz \& Smith \cite{Katz88},
Raubenheimer and Smit \cite{Rau97}), can be excluded from the list of
reasonable options. At the same time, the $\gamma$-ray source must be
genetically connected to the neutron star. This requirement follows
{\it not} from the effect of $\gamma$-ray modulations with the pulsar spin,
which would still need additional confirmation by future
observations, but from the general considerations of the source energetics.

Indeed, the radiation flux detected by EGRET during 2 weeks in 1994
corresponds to an average source luminosity of
$L_\gamma (100\, {\rm MeV} \leq E \leq 10\,{\rm GeV})
\simeq 5\times 10^{36} \,\rm erg s^{-1}$
(Vestrand et al. \cite{V97}), and the mean flux detected in
1997-1999 by the Mark-6 telescope corresponds to a
luminosity  $L_{\gamma}(>400\,\rm GeV) \sim 10^{36} \, erg s^{-1}$. For the
parent relativistic particles, either electrons or protons, these
luminosities inevitably imply an acceleration power $P_{\rm acc} \sim
10^{37}\,\rm erg s^{-1}$ or higher. The optical/UV luminosity of the
Krzeminski star corresponding to the bolometric magnitude $M_{\rm bol} \simeq
-9$ (Hutchings et al. \cite{Hut79}) is really high, $L_{\rm bol} \simeq
10^{39}\,\rm erg/s $\footnote{Note that for a massive star with $M\sim 20
\,\rm M_{\odot}$ this is already comparable with, but still significantly
below the Eddington luminosity
$L_{\rm Edd}=1.3\times 10^{38} (M/M_{\odot} )\,\rm erg s^{-1}$.}.
It would suffice if a small fraction of this luminosity could somehow
be converted to the acceleration of relativistic particles.
In principle, this might be due to a radiatively driven supersonic
stellar wind and subsequent production of shocks in such winds. Meanwhile,
for the characteristic speed of the stellar wind $v \sim 10^{3} \,\rm km
s^{-1}$ and the mass-loss rate $\dot{M} \sim 10^{-6} M_{\odot} \rm yr^{-1}$
(Clark et al. \cite{Clark88}) the entire kinetic energy of the wind produced
by the Krzeminski star makes only
$P_{\rm wind}= \dot{M} v^2/2 \simeq 3\times 10^{35} \,\rm erg s^{-1}$.

Any reasonable possibility for production of a kinetic power exceeding
$10^{37}\,\rm erg s^{-1}$ in Cen X-3/V779 Cen binary to be is therefore
to be attributed to the neutron star. For the 4.8\,\rm s spin period of Cen
X-3 the classical mechanism of the magnetic dipole radiation falls several
orders of magnitude below the values required. Therefore the only remaining
principal option for the prime energy source for the acceleration of
relativistic particles in this system remains the kinetic energy of the inner
accretion disc formed around the neutron star. Most of the kinetic
energy of the disc ends up on the neutron star surface in the form
of the thermal plasma energy responsible for the observed X-rays with the
luminosity reaching (in the {\it high} state) $L_{\rm X } \geq  10^{38}\,\rm
erg \, s^{-1}$. It is possible, however, that a significant fraction of
the accretion disc energy can be ejected from the system producing powerful
outflows in the form of two-sided jets observed at least in some
class of close X-ray binaries currently identified as `microquasars'
(see Mirabel \& Rodriguez 1999 for a review), which may contain either a
stellar-mass black hole or a
neutron star. A spectacular example of a microquasar containing an
accreting neutron star that produces powerful subrelativistic ($v_{\rm jet}
\approx 0.26 c$) jets is SS 433, with the jet kinetic energy
estimated in the range from
$\sim 10^{39} \,\rm erg s^{-1}$ to $4\times 10^{40} \,\rm erg s^{-1}$
(see Margon \& Anderson \cite{Margon89}, Panferov \cite{Pan99}). In this
regard it is also worth noting the recent report by Jernigan et al.
(\cite{Jer00}) of the detection of high-frequency QPOs (`quasi-periodic
oscillations') in the X-ray emission of Cen X-3, which is a characteristic
feature of microquasars.

The principal scenarios for $\gamma$-ray production in Cen X-3 are thus
connected with powerful jets of plasma driven by the inner
accretion disc. In one such scenario these jets would
create strong shocks while propagating through the rather dense wind driven
by the O-star. Subsequent acceleration of particles on these shocks could
result in the creation of a $\gamma$-ray source around the region of
jet propagation (or its damping) on large spatial scales $l_{\rm s}$
comparable with, or more probably significantly, exceeding the size of the
binary $\sim 10^{12}\,\rm cm$. Obviously, such a spatially extended model
will not be able to explain episodes of pulsed $\gamma$-ray emission that may
have been detected from the system. Evidence for pulsed emission from Cen~X-3
is far from conclusive, but if it is there it probably constitutes only a
fraction of the overall $\gamma$-radiation of Cen X-3. An interpretation of
the `anomalously' pulsed $\gamma$-ray episodes may be possible in the
framework of compact-source model scenarios that would suppose direct
interaction of the jet either with material ejected from the inner parts of
the accretion disc, or with a dense external target that might accidentally
fall under the jet. Below we will discuss these model possibilities.

\subsection{Spatially extended source model}

In the framework of this model that assumes acceleration of
particles on the shocks driven into the supersonic (and therefore
favourable for generation of shocks) wind of the O-star by
jets appearing from the inner accretion disc, the source size
$l_{\rm s} \gg 10^{12}\,\rm cm$, so that the $\gamma$-ray source is
surrounding the X-ray pulsar. This circumstance allows us to
exclude with a significant confidence the hadronic ($\pi^{0}$-decay)
origin of $\gamma$-rays for the spatially extended model.

Indeed, the energy loss time of protons due to inelastic collisions with
ambient gas with a density  $n_{\rm gas}$ (in terms of nucleons, or
`H-atoms') is
\begin{equation}
 t_{\rm pp} =  ( K_{\rm p} \sigma_{\rm pp} n_{\rm gas} c)^{-1}
\simeq 2.5\times 10^{5} \left( \frac{n_{\rm gas}}{10^{10} \,\rm cm^{-3}}
\right)^{-1}\; \rm  s \,
\end{equation}
where $K_{\rm p}\approx 0.45$ is the inelasticity coefficient and
$\sigma_{\rm pp} \sim 30 \,\rm mb$ is the cross-section of nuclear
interactions of protons at GeV energies which slowly increases with
the particle energy. Taking then into account also
that in $pp$-collisions only $\simeq 1/3 $ of the initial energy of
relativistic particles is eventually transformed to $\gamma$-rays (the rest
goes to the secondary electrons and neutrinos), the total energy $W_{\rm p}$
of relativistic protons which is needed for $\gamma$-ray production with
luminosity $L_{\gamma}$ is estimated as
\begin{equation}
W_{\rm p} \simeq 3 t_{\rm pp} \times L_{\gamma} \;
\end{equation}

For a $\gamma$-ray source extending up to characteristic distances
$R_{\rm s}$ measured from the centre of the O-star, the source volume can be
generally written as
$V_{\rm s} =f_{\rm V} \times 4 \pi R_{\rm s}^3 /3 $
where $f_{\rm V}\leq 1$ is the volume filling factor. The energy density of
relativistic protons $w_{\rm p} = W_{\rm p} / V_{\rm s}$ should then be
compared with the energy density
$w_{\rm gr} = \frac{G M m_{\rm p} n_{\rm gas}}{R_{\rm s}}$ of the thermal
gas in the
gravitational field of the massive companion star with $M\simeq 20 M_{\odot}$
which defines the gravitational confinement potential of the binary.
Obviously, in any reasonable scenario the energy density of relativistic
charged particles that can accumulate ({\it be frozen}) in the ambient
plasma should not exceed the gravitational field energy density, otherwise
the relativistic particle gas cannot be confined to a scale $R_{\rm s}$ but
would act to inflate the source, also strongly affecting the gas dynamics in
the binary system. Then from the condition $w_{\rm p} \leq w_{\rm gr}$ we can
estimate the maximum $\gamma$-ray luminosity $L_{\pi^0}^{\rm max}$ that can
be produced by relativistic  protons:
\begin{eqnarray} L_{\pi^0}^{\rm max} & = & \frac{4 \pi}{9} f_{\rm V} G M
                                           m_{\rm p} K_{\rm p}
\sigma_{\rm pp} N_{\rm H}^2 c \\
 & \simeq &  2.5 \times 10^{34} \,f_{\rm V}
\left( \frac{M}{20 \, M_{\odot}} \right)
\left( \frac{N_{\rm H}}{ 10^{23}\,\rm cm^{-2} }\right)^{2} \,
\rm erg s^{-1} \; \nonumber
\end{eqnarray}
Here $N_{\rm H}= n_{\rm gas} \times R_{\rm s}$ is the gas column density of
the $\gamma$-ray source which we have normalized to $10^{23}\,\rm cm^{-2}$.
This is a reasonable maximum value for $N_{\rm H}$ that corresponds to the
gas density $n_{\rm gas}\leq 10^{11}\,\rm cm^{-3}$ of the wind at distances
$R_{\rm s} \geq 10^{12} \,\rm cm$ (Clark et al. \cite{Clark88}). Eq.(3) shows
that the maximum luminosity to be expected in $\gamma$-rays of hadronic
origin is orders of magnitude below the $\gamma$-ray luminosities detected
from Cen~X-3.

It should be noted that Eq.(3) is actually a rather universal upper limit
for the $\pi^0$-decay $\gamma$-ray luminosity that can be produced
({\it quasi-stationarily}) by a source gravitationally confined to a central
mass $M$. In particular, this relation is valid for $\pi^0$-decay
$\gamma$-rays that can be produced in the accretion discs of neutron stars or
black holes ({\it including} supermassive BH in AGNs). For Cen X-3  we should
substitute $M\rightarrow M_{\rm n}\simeq 1.2\,  M_{\odot}$ and take into
account that for any reasonable disc geometry $f_{\rm V}$ cannot exceed $\sim
0.1$. Then even assuming $N_{\rm H} \simeq  8\times 10^{23}\,\rm cm^{-2}$
which is the characteristic maximum column density accumulated across the
accretion disc of Cen X-3, as deduced by (Nagase et al. \cite{Nagase92}) from
the X-ray line variations in the pre-eclipse dips, we conclude that any
hadronic model invoking the accretion disc as a site for $\gamma$-ray
production would be far too inefficient for an explanation of the observed
fluxes, or otherwise the relativistic proton gas would blow up the accretion
disc itself.

Considering now models that assume a leptonic origin of $\gamma$-rays,
it becomes crucial for the existence of a quasi-stationary large-scale source
that the energy loss time of the electrons in Cen X-3 is much shorter than
$t_{\rm pp}$. At distances $R \geq 1.5 R_{\rm O}$ from the O-star the energy
density of  the thermal optical/UV radiation field is  $w_{\rm UV}= L_{\rm
bol}/4\pi R^2 c$, therefore the cooling time of an electron with
Lorenz-factor $\gamma$ due to inverse Compton (IC)  radiation can be
estimated as
\begin{equation}
t_{\rm IC} \simeq 1.1\times 10^{4} (R/10^{12}\,\rm cm)^2 \gamma^{-1} \;
\rm s \; .
\end{equation}
This estimate is valid in the Thompson limit for IC scattering when the
parameter $b=4 \gamma (\epsilon_0/m_{\rm e} c^2) < 1$. For the O-star in Cen
X-3 with $T\sim 35\, 000\, \rm K$ the mean energy of the thermal photons is
$\epsilon_0 \sim 3\, k T \simeq 9 \,\rm eV$, therefore the transition from
the Thompson limit to the relativistic Klein-Nishina limit of IC scattering
occurs at $\gamma \sim \gamma_{\rm KN} =1.4 \times 10^4$. Thus, for electrons
with $\gamma \sim 10^3-10^4$ responsible for production of high-energy
$\gamma$-rays the IC cooling time is orders of magnitude smaller than the
cooling time of relativistic protons   \footnote{Note that with increasing
$R$ the gas density $n_{\rm gas} $ drops approximately as $\propto R^{-2}$,
so $t_{\rm pp} \propto R^2$ similar to $t_{\rm IC}$.} given by Eq.(1). For
electrons with $\gamma \geq \gamma_{\rm KN}$ the IC loss time does not drop
further as Eq.(4) would predict, but rather increases, so that for VHE
electrons the IC cooling time at distances
$R$ is about $(10^2-10^3) \times (R/10^{12}\,\rm cm)^2$. Thus, for these
electrons as well the case is still $t_{\rm IC} \ll t_{\rm pp}$, which leads
to a total energy $W_{\rm e}$ needed in relativistic electrons which is much
smaller than $W_{\rm p}$.

Thus, a spatially extended $\gamma$-ray source assumes diffusive shock
acceleration (e.g. see Drury 1983, Blandford \& Eichler 1987) of the
electrons on spatial scales $R_{\rm s}$ comparable or exceeding the
characteristic length scale of the binary. In order to estimate a plausible
size  $R_{\rm s}$, let us  estimate the characteristic time needed for
acceleration up to energies $E \sim 10\,\rm TeV$. Using the results of
Lagage \& Cesarski \cite{Lag83}, in the Bohm diffusion limit this time can
be expressed as
\begin{equation}
t_{\rm acc} \simeq 7\times 10^4 \frac{E}{10\,\rm TeV}
\left( \frac{B}{1\,\rm G} \right)^{-1} \left( \frac{v_{\rm sh}}{3000\,\rm
km s^{-1}} \right)^{-2} \rm s \, ,
\end{equation}
where $v_{\rm sh}$ is a mean shock speed. Since the characteristic speed of
the stellar wind in Cen X-3 is $v_{\rm w}\sim 10^3\,\rm km s^{-1} $, the
normalization of $v_{\rm sh}$ to $\sim 3000 \,\rm km s^{-1}$ used above seems
reasonable. Now taking into account that because of advection in the plasma
wind all relativistic particles leave the radial scales $R$ at least on the
convective escape timescale $t_{\rm esc}\simeq R/v_{\rm w}$, the
characteristic size of the region required in this scenario for production of
multi-TeV electrons should satisfy the condition $t_{\rm acc}(10\,\rm TeV)
\leq t_{\rm esc}$. As it follows from our calculations below, the magnetic
fields of order $B \sim 1 \, \rm G$ in the stellar wind of Cen X-3 would not
contradict the fact of non-detection of radio fluxes from Cen X-3 (see
Vestrand et al. \cite{V97}). Therefore, from the condition  $t_{\rm esc}\geq
t_{\rm acc}$, the size of the TeV $\gamma$-ray source is estimated as $R_{\rm
s}\geq 5\times 10^{12}\, \rm cm$. On the other hand, the source size should
not significantly exceed $10^{14} \,\rm cm$ because of the  fast decline of
the density of the UV photon field from the O-star, which essentially reduces
the efficiency of the IC $\gamma$-ray production  (as $t_{\rm IC} > t_{\rm
esc}$).

Note that $R_{\rm s}$ is much larger than the radius $R_{\rm O} =8.6
\times 10^{11} \,\rm cm$ of the Krzeminski star or the distance $a=1.3\times
10^{12} \,\rm cm$ between the stars. This implies a source volume filling
factor $f_{\rm V} \sim1$. More importantly, the large $R_{\rm s}$ suggests
that the $\gamma$-ray source is {\it quasi-stationary} on time scales
$\geq 1\,\rm day$.

Provided that the volume occupied by the shock fronts (i.e. where the
particle acceleration occurs) is significantly smaller than the $\gamma$-ray
source entire volume, one can describe the energy distribution of
relativistic electrons $N(E)$ by a kinetic equation (see e.g. Ginzburg \&
Syrovatskii 1964):
\begin{equation}
\frac{\partial N}{\partial t} = \frac{\partial}{\partial E}
 \left( P\, N \right) - \frac{N}{t_{\rm esc}} + Q \, ,
\end{equation}
where $Q\equiv Q(E,t)$ is the rate of the electron acceleration/injection,
and $P = -{\rm d} E/{\rm d}t$ corresponds to overall electron energy losses
averaged over the volume of the source. The injection spectrum of
the electrons is approximated as
\begin{equation}
Q(E) \propto E^{-\alpha_{\rm e}} \, \exp{(-E/E_0)} \, ,
\end{equation}
with $\alpha_{\rm e} \sim 2$ typical for the diffusive acceleration on the
strong shock fronts, and $E_0 \geq 10 \, \rm TeV$ for the exponential cutoff
energy.

\begin{figure}[tbp]
\centerline{\epsfig{file=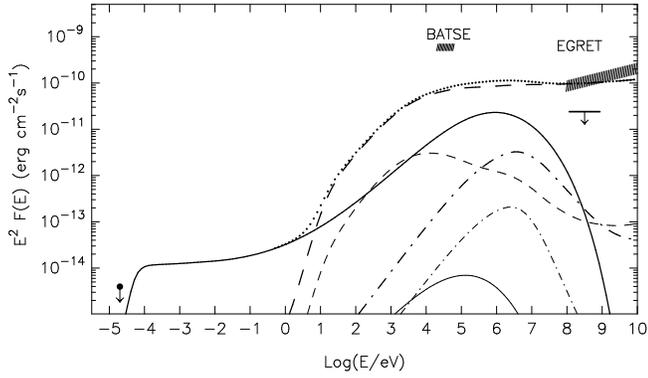, width=8.5cm}}
\caption{The spectra of synchrotron (solid lines), bremsstrahlung (dot-dashed
lines), and inverse Compton (dashed lines) radiations expected when
calculated in the framework of a quasi-stationary model with $R_{\rm s} =
6\times 10^{12}  \,\rm s$, assuming a magnetic field $B=0.7\,\rm G$, and
injection spectrum of electrons with $\alpha_{\rm e} = 2.0$, $E_{0}=15\,\rm
TeV$, and $P_{\rm acc}= 2\times 10^{37}\,\rm erg$. Radiation fluxes produced
outside the source (zone 2, see text) are shown by thin lines.}
\label{Fig:Arm1}
\end{figure}

\begin{figure}[htbp]
\centerline{\epsfig{file=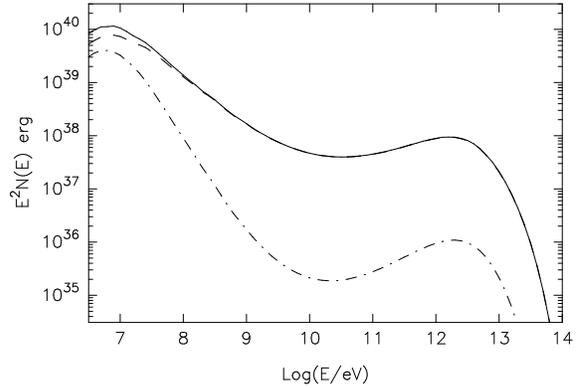, width=7.5cm}}
\caption{The energy spectra of relativistic electrons formed in zones 1
(dashed) and zone 2 (dot-dashed), and the total spectrum (solid line)
for the extended source model with parameters in Fig.~\ref{Fig:Arm1}}
\label{Fig:Arm2}
\end{figure}

In Figure \ref{Fig:Arm1} we show the spectra of non-thermal radiation from
radio wavelengths up to the HE $\gamma$-ray band calculated for such
a quasi-stationary spatially extended source with $R_{\rm s} = 6\times
10^{12} \,\rm s$, assuming injection of relativistic electrons with an
overall power $P_{\rm acc} = 2\times 10^{37}\,\rm erg$. In order to assess a
contribution to the overall fluxes due to electrons taken away to large
distances with the stellar wind, the calculations are done in a two-zone
model approach (see Atoyan et al. \cite{Atoyan00} for details), which assumes
that there is no acceleration of fresh electrons at $R > R_{\rm s}$ (zone 2),
but there is rather an injection of electrons escaping from the main source
(zone 1) at the rate $Q^{\prime}= N(E)/t_{\rm esc}$. In Figure \ref{Fig:Arm1}
the fluxes of synchrotron, bremsstrahlung and IC radiation produced
in zone 2 are shown by thin lines. Obviously these fluxes are
negligible compared with the respective ones produced in the region $R\leq
R_{s}$ where the effective electron acceleration is supposed. It tells us
that for most of the electrons the characteristic cooling time inside the
source is indeed significantly shorter than their escape time. This is
apparent also in Figure \ref{Fig:Arm2} where the energy distribution of
relativistic electrons in zone 2 is shown by a dot-dashed line. Note an
unusual profoundly concave shape of the electron distribution which is due to
the Klein-Nishina effect in the IC energy losses of VHE
electrons in the thermal radiation field of the O-star.

The full dot in Figure \ref{Fig:Arm1} corresponds to the upper flux limit
$S_\nu \leq 70 \,\rm mJy$ at 5 GHz in the direction of Cen X-3 (Vestrand et
al. \cite{V97}). The heavy solid curve in this figure shows that due to
synchrotron self-absorption at frequencies below 10-15 GHz the
values of the magnetic field up to $\sim 1 \,\rm G$ at distances $\leq
10^{13} \,\rm cm$ can still be accepted. However an assumption of
a stronger magnetic field would result in significant synchrotron
fluxes at sub-mm wavelengths which would then already be observed.

\begin{figure}[tbp]
\centerline{\epsfig{file=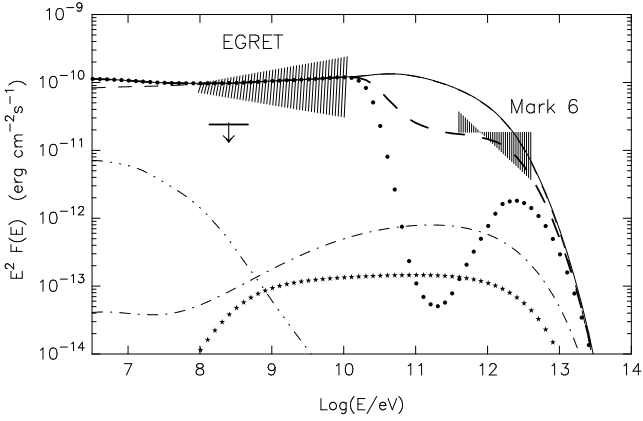, width=8.5cm}}
\caption{The spectra of IC $\gamma$-rays calculated for the spatially
extended source with parameters in Fig.~\ref{Fig:Arm1}. Contributions to the
total unabsorbed IC radiation (solid line) due to upscattering of the UV
radiation of the companion O-star (thin dashed line; distinguished only at
$E < 10^8\,\rm eV$), X-ray pulsar photons (3-dot--dashed line) and
synchrotron photons (dot-dashed line) are shown. The heavy dashed line shows
the spectrum escaping the source. The heavy dotted line shows the fluxes that
would be expected if the same unabsorbed radiation were produced at the
pulsar orbit; the stars show the unabsorbed radiation of hadronic origin
(see text). The hatched region at TeV energies corresponds to the average
flux detected by the Mark-6 telescope during 1997 (which is higher than the
mean for 1997-1999, but {\it closer} to 1994) calculated for differential
power-law indices between $\alpha_{\gamma}=2$ and 3. The range of
differential fluxes, from median to low, detected by EGRET during October
1994, as well as the upper flux limit for earlier observation period, are
also shown.}
\label{Fig:Arm3}
\end{figure}

In Figure \ref{Fig:Arm1} we also show a typical level of X-ray fluxes
detected by BATSE from Cen X-3 in the active (`high') state. These fluxes are
predominantly of thermal origin, and much higher than the non-thermal
fluxes. Note however that the latter could, in principle, show up in the
hard X-ray/soft $\gamma$-ray band, $E \sim (0.1-10) \,\rm MeV$, where the
thermal radiation of the neutron star drops. In the region of HE
$\gamma$-rays the IC radiation flux $\alpha \simeq 2$ is somewhat steeper
than the October 1994 flux of Cen X-3 with a hard {\it mean} differential
power-law index $\alpha_{\rm obs}= 1.81$ detected by EGRET (hatched region in
Figure \ref{Fig:Arm1}). However, if we take into account the reported
uncertainty in the power-law index $\pm \Delta \alpha = 0.37$, the model
appears in good agreement with the EGRET data (see Figure
\ref{Fig:Arm3}, where the upper boundary of the hatched zone at 0.1-10 GeV
energies shows the median flux, while the lower one corresponds to
$\alpha_{\rm obs}= 2.18$).

In Fig.~\ref{Fig:Arm3} IC $\gamma$-ray fluxes extending beyond 10 TeV are
presented. The solid curve corresponds to the {\em unabsorbed} flux produced
in the source. The principal target photons for the IC scattering of the
electrons in this model is the thermal UV radiation of the O-star,
the contributions due to X-ray (3-dot--dashed) and synchrotron
(dot-dashed) photons being negligible at all $E$. The heavy dashed line shows
the fluxes of $\gamma$-rays escaping the source, after taking into account
their absorption in the UV photon field. Calculations assume that
$\gamma$-rays are produced uniformly at the scales $R\leq R_{\rm s}$.

In order to demonstrate a possible impact of photoabsorption, in Figure
\ref{Fig:Arm3} we show by full dots the flux that would escape the
binary if the same unabsorbed $\gamma$-radiation were to be produced
close to the neutron star in the orbital phase $\phi \sim \pm 0.25$. In that
case the optical depth $\tau_{\gamma \gamma} (E)$ would reach a maximum
$\simeq 7$ at energies $E\sim 200\,\rm GeV$, and $\gamma \gamma$-absorption
would remain very significant in the entire region from
$E \geq \,\rm 30 GeV $. Because $\tau_{\gamma \gamma}$ depends strongly on
$\phi$, essentially becoming less than 1 at a phase of $\phi \sim 0.5$, a
strong orbital modulation of VHE $\gamma$-ray fluxes produced close to the
neutron star would be expected. In our calculations we did not take into
account a possible contribution connected with the development of pair-photon
cascades in the atmosphere of Cen X-3, which in practice remains small and
cannot in any way significantly compensate the impact of absorption on the
spectra of VHE $\gamma$-rays (see Bednarek \cite{B00}).

The total amount of energy accumulated in relativistic electrons shown in
Figure \ref{Fig:Arm2} is $W_{\rm e} = 2.4\times 10^{40} \,\rm erg$. For
comparison of the efficiencies of hadronic and leptonic models,
in Figure \ref{Fig:Arm3} we show by stars the fluxes of $\pi^0$-decay
$\gamma$-rays produced by relativistic protons with the same amount of total
energy, $W_{\rm p}= 2.4\times 10^{40} \,\rm erg$.

It should nevertheless be said that the energy $W_{\rm e}$ is still
significantly larger than the total potential energy $W_{\rm G}$ of the
plasma in the gravitational field of the binary at scales $\leq 10^{13}\,\rm
cm$. The latter can be expressed directly through the mass loss rate
$\dot{M}= 4\pi m_{\rm p} n_{\rm gas} R^2 v_{\rm w}$
of the O-star if we approximate the stellar wind speed as a constant,
$v_{\rm w}(R) \simeq 10^{3}\,\rm km s^{-1}$:
\begin{equation}
W_{\rm G} \simeq 2\times 10^{39}\, \dot{M}_{-6} \,
\frac{M_{\rm O}}{20\,M_{\odot}} \, \left(
\frac{v_{\rm w}}{10^{3} \, \rm  km s^{-1}  } \right)^{-1}
\; \rm erg \, ,
\end{equation}
where $ \dot{M}_{-6} \equiv \dot{M}/10^{-6} \,M_{\odot} \rm yr^{-1} $.
However, $W_{\rm e}$ is comparable with the total kinetic energy of the wind
\begin{equation}
W_{\rm kin} \simeq 3.2 \times 10^{40} \, \dot{M}_{-6} \,
\frac{v_{\rm w}}{10^{3} \, \rm  km s^{-1}  } \,
\frac{R}{10^{13}\,\rm cm} \; \rm erg \,.
\end{equation}
It then implies that this model can still be magnetohydrodynamically
self-consistent because the pressure of the relativistic electron gas
can be `balanced' by the kinetic ({\it ram}) pressure. Obviously, the
electrons in these circumstances should contribute very significantly to the
gas dynamics of the stellar wind.

It is worth comparing here requirements that would follow from similar
arguments in the case of a spatially extended hadronic model. In order to
explain the observations one would require an overall energy in relativistic
protons almost 3 orders of magnitude larger than the flux of gamma-rays of
hadronic origin (shown by stars in Fig. \ref{Fig:Arm3}); i.e.
$W_{\rm p} $ would exceed $10^{43}\,\rm erg$. Such a huge amount of
energy at a scale of $R_{\rm s} = \leq  10^{13} \,\rm cm$, as assumed for Fig.
\ref{Fig:Arm3}, is very problematic to expect from a binary star
generally, and is far beyond the gravitational confinement abilities of the star,
as discussed above. Moreover, Eq.(9) predicts that one would then expect such a
relativistic proton gas would drive a wind in Cen.~X-3 moving at
subrelativistic speeds ($v_{\rm w}$). Assuming a larger source size
would not help, as might be hinted by the same Eq.(9), on the
contrary it will make the situation even worse because of a fast drop off of the
mean gas density at larger distances resulting in a fast increase of the
overall energy of relativistic protons needed. Only a rather compact source that
could provide gas densities much higher than are found in the 0-star
wind, so resulting in a reasonable $W_{\rm p}$, giving a reasonable
chance to provide the observed fluxes (see below). Note, however, that this
does not mean that in the framework of a large-scale model protons are not
allowed to be {\it accelerated} with an efficiency comparable with
the electrons.

The relativistic electron gas pressure in the source can be
reduced further if we take into account that most of the energy in
Figure \ref{Fig:Arm2} is due to particles with $E\leq 100 \,\rm MeV $ which
do not contribute to production of HE $\gamma$-rays. Then we can reduce
$W_{\rm e}$ assuming a very hard spectrum of accelerated electrons
with $\alpha_{\rm e} < 2$. Note that taking into account non-linear
modification effects at the strong shock fronts, power-law spectral indices
of the shock accelerated particles even as hard as $\alpha_{\rm e} =1.5$
could be expected (Malkov \cite{Malkov97}).

\begin{figure}[tbp]
\centerline{\epsfig{file=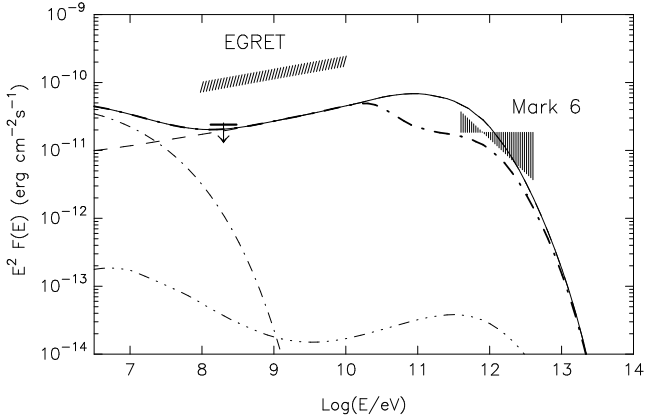, width=8.5cm}}
\caption{Gamma-ray fluxes (heavy dot-dashed line) expected in the framework 
of a quasi-stationary source with 
$R_{\rm s}=10^{13}\,\rm cm$,  $B=1\,\rm G$ in the case of a hard
power-law spectrum of accelerated electrons with $\alpha_{\rm e}=1.75$, and
total electron acceleration power $P_{\rm acc}=6\times 10^{36}\,\rm erg
s^{-1}$. Contributions due to IC (dashed line), synchrotron (thin dot-dashed
line) and bremsstrahlung (3-dot--dashed line) radiation mechanisms are
shown.}  
\label{Fig:Arm4} 
\end{figure}

In Fig.~\ref{Fig:Arm4} we show the $\gamma$-ray spectra calculated for 
$\alpha_{\rm e} = 1.75$, and assuming an injection rate of electrons with
$P_{\rm acc}=6\times 10^{36}\,\rm erg s^{-1}$. In this case the total 
energy of the electrons in the region $R\leq R_{\rm s}=10^{13}\,\rm cm$ is
only $W_{\rm e}=4.1\times 10^{39}\,\rm erg$. Note that in the HE 
$\gamma$-ray domain the {\it shape} of the model spectrum
now agrees well with the median spectral index of the fluxes detected
by EGRET in October 1994. Meanwhile, the absolute fluxes of HE
$\gamma$-radiation are now in agreement with the upper flux limits reported
for the 9 month earlier observing session of Cen X-3 by EGRET
(Vestrand et al. \cite{V97}). For calculations in Fig.~\ref{Fig:Arm4}
we have chosen rather a small acceleration rate of
electrons, $P_{\rm acc}=6\times 10^{36}\,\rm erg s^{-1}$, in order to
demonstrate that in principle in the framework of this quasi-stationary model
with $\alpha_{\rm e} < 2 $ it is possible to expect episodes (on time scales
$\geq 1\,\rm d$) of $\gamma$-radiation that could have been detected in the
VHE $\gamma$-rays, but would be missed by EGRET.     

In conclusion, a spatially extended quasi-stationary source model seems
capable of providing a phenomenologically self-consistent explanation for
most of the currently existing data of $\gamma$-ray observations of Cen X-3.
The single observational feature which cannot be addressed by this model is
an interpretation of possible modulations of the $\gamma$-ray emission with
the pulsar spin period $P_0$. If genuine, such modulations would either
require an alternative or an additional much more compact $\gamma$-ray
source.          
 
\subsection{Compact source models}  

An obvious requirement on any model that would be able to explain a
$\gamma$-ray emission modulated at the pulsar spin period $P_0$ is that the
effective source radius $R_{\rm s}$ has to be noticeably smaller than 
$c P_{0}/2$. For Cen X-3 therefore $R_{\rm s}$ should be $\leq 5\times
10^{10}\,\rm cm$. Another requirement is that the energy loss time 
$t_{\rm loss}$ of relativistic particles producing  
pulsed $\gamma$-rays should also be smaller than $P_0$. Then for 
$t_{\rm loss} \sim 1\,\rm s$ the energy density of relativistic particles can
be estimated as $ w_{\rm rel} \geq 10^{4} \,\rm erg/cm^3$ using the relation
$W_{\rm rel}\geq L_{\gamma} t_{\rm loss}$ for the total particle energy in
the source. Such high energy densities can be reached in the binary only in
the inner accretion disc around the neutron star, which however
cannot be a site for the $\gamma$-ray emission because of the absence of any
indications for the orbital modulation of the fluxes in both the HE
and VHE domains, as discussed in Section 2 above. This implies that
the source of pulsed $\gamma$-radiation in Cen X-3 cannot be confined, 
so that generally one would expect only rather short episodes of pulsed
$\gamma$-ray emission.  
  
As we have discussed above, a source of the observed energetic  
gamma-radiation in this X-ray binary should be connected with jets generated
by the accretion disc around the neutron star. There are then two principal
types of model, leptonic and hadronic, for the production of episodic pulsed
radiation, both of which would assume a compact source (`clouds', `blobs')
propagating in the jet. Pulsations could be produced if a powerful
relativistic energy outflow, in the form of either electromagnetic Poynting
flux or a beam of relativistic particles, modulated with the pulsar spin
period propagates in the jet region and accelerates/injects relativistic
particles in the source.   

A hadronic {\it `beam-target'} model for production of pulsed VHE 
$\gamma$-radiation in X-ray binaries has been suggested earlier by Aharonian
\& Atoyan (\cite{AA91,AA96}). This model assumes that a powerful beam of
relativistic protons accelerated in the vicinity of the pulsar hits a dense
plasma cloud that may appear in the jet propagation region. As it follows
from Eq.(1), the gas density in the cloud should be very high, $n_{\rm gas}
\geq 10^{15} \,\rm cm^{-3}$, in order to provide fast energy losses
for the protons, $t_{\rm pp} < P_0 =4.8 \,\rm s$. For a cloud with a radius
$R_{\rm cl}\sim 10^{10}\,\rm cm$ this implies a cloud mass 
$M_{\rm cl} \geq 10^{22} \,\rm g$. 
One may phenomenologically suppose that either such dense plasma blobs are
ejected from the massive optical companion star, and then they may cross a
conical region of jet propagation. We may also speculate that compact dense
plasma blobs can be  ejected from the accretion disc or be created in the jet
region, and then the `beam-target' interaction takes place in the episodes
when relativistic protons are sporadically accelerated in the central engine
(e.g. by the dynamo action, magnetic field line reconnection, etc) forming
relativistic proton beams streaming in the jets.  
this regard that very dense and compact ($\sim 10^8 \,\rm cm$) gas clouds
responsible for the observed optical emission in the powerful relativistic
jets of the prominent binary SS 433, are supposed to be produced/condensed
due to thermal  instabilities in those jets (see Brinkmann et al.
\cite{Brink88}, Panferov \cite{Pan99}).   

A relativistic proton beam interacting with the gas in the cloud would
produce $\pi^0$-decay $\gamma$-rays. Non-thermal radiation in this
model will be contributed also by secondary electrons ($e^{\pm} $). At the
same time some fraction (a few per cent, see Aharonian \&
Atoyan \cite{AA91}) of the injected proton energy would inevitably go to
Coulomb (`ionization') losses heating the cloud up to temperatures  $T_{\rm
cl} \sim (5-10)\times 10^4\,\rm K$. This results in a very high density UV
radiation field in the cloud, so that its opacity with respect to VHE
$\gamma$-rays can be very high. The spectra of escaping $\gamma$-rays are
then essentially defined by the pair-photon cascade developing in the cloud.
At the same time, because of the high pressure created, the source would
expand with a speed presumably of order of the sound speed, $\sim \sqrt{k
T_{\rm cl}/m_{\rm p}}$, if not confined by some external pressure. Therefore
the radiation spectra predicted in this expanding source should be rapidly
evolving. 

\begin{figure*}
\centerline{\epsfig{file=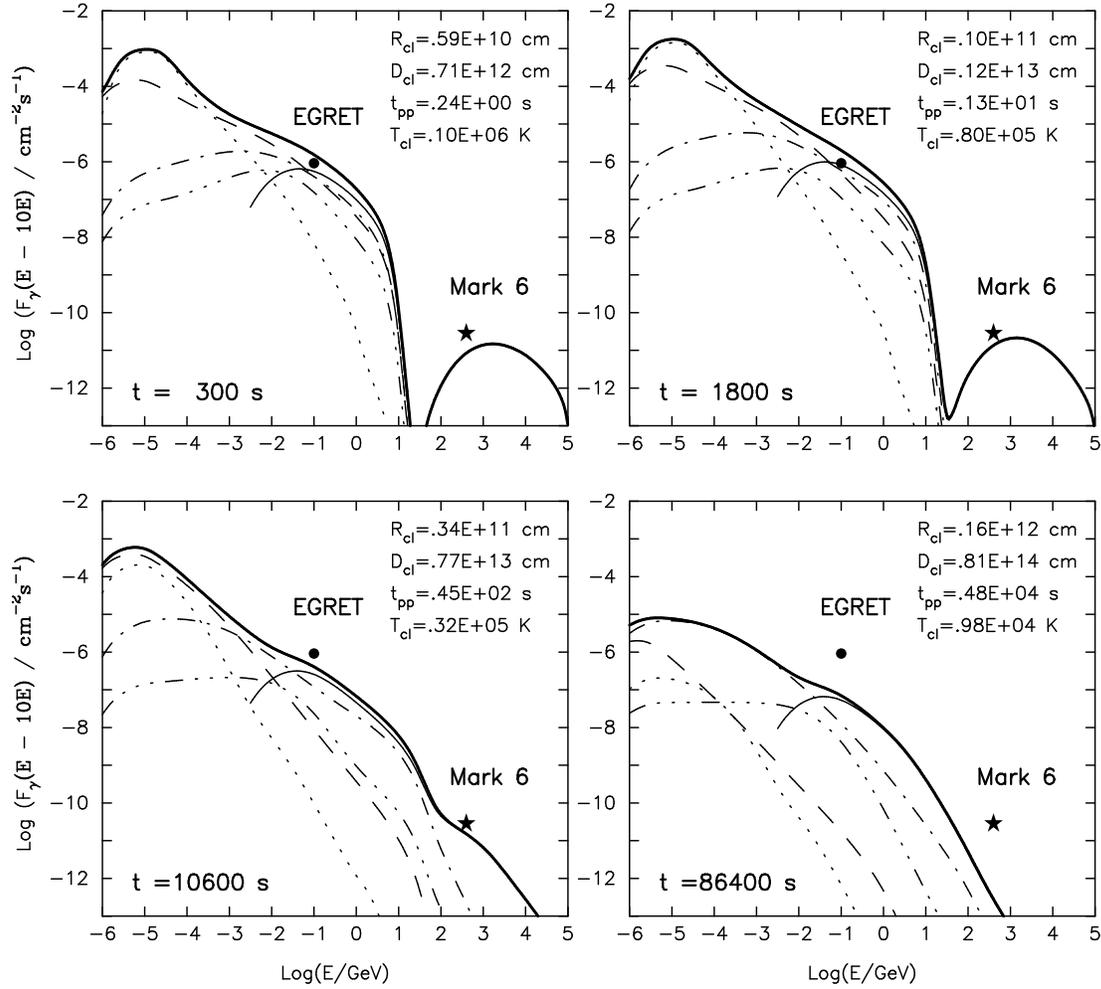, width=14.5cm}}
\caption[]{The radiation spectra expected at different times from a
compact heavy target with $M_{\rm cl}=10^{22}\,\rm g$ falling under a 
powerful beam of relativistic protons. The fluxes and principal parameters of
the cloud, as the radius $R_{cl}$, the distance $D_{\rm cl}$ from the pulsar,
the $p-p$ interaction time $t_{\rm pp}$, and the cloud temperature  $T_{\rm
cl}$, are shown. The contributions of different radiation processes are shown
by solid line ($\pi^0$ decays), dashed line (synchrotron radiation),
dot-dashed line (inverse Compton), 3-dot--dashed line (electron
bremsstrahlung), dotted line ('Comptonization' - the multiple Compton
scattered radiation in the thermal gas). The heavy solid line corresponds to
the total spectrum escaping the cloud. Note that the integral fluxes here are
integrated 'per decade of the energy interval', which allows to
better preserve the information about differential radiation flux as
well.}
\label{Fig:Arm5} 
\end{figure*}

In Fig.~\ref{Fig:Arm5} we show the non-thermal radiation spectra expected in
this model at 4 different times $t$, from 5 min to 1\,day ($=86400\,\rm s$), 
after a cloud with a mass $M_{\rm cl} =10^{22}\,\rm g$ and an initial size
$R_{\rm cl}=5\times 10^{9} \,\rm cm$ is hit by a powerful beam of
relativistic protons at the distance $D_{\rm cl} = 8\times 10^{11} \,\rm cm$
from the neutron star. The method of calculations used and other
details of the model are found in Aharonian \& Atoyan (1996). The spatial
angular density of the beam power needed in this model is rather high, 
${\rm d} P_{\rm beam}/{\rm d}\Omega = 3\times 10^{41}\,\rm erg \, s^{-1}
sr^{-1}$ in Fig.~\ref{Fig:Arm5}. This implies that in order to have
reasonable total energetics, the relativistic proton beam should be
collimated to within a few degrees. Thus, for the collimation angle $5^\circ$
the total beam power would be $P_{\rm beam}=1.8\times 10^{39}\,\rm erg
s^{-1}$. We remind the reader in this regard that the collimation angle of
SS~433 jets is $\theta_{\rm j}\leq 1.4\,^{\circ}$, and the jet kinetic energy
is estimated up to several times $10^{40}\,\rm erg s^{-1}$ (e.g. Margon \&
Anderson \cite{Margon89}).

Under the pressure of such a powerful beam the expanding clouds would be 
expelled from the binary reaching speeds of order $10^{4}\,\rm km s^{-1}$.
This is seen in Fig.~\ref{Fig:Arm5} where we show the increasing distance
$D_{\rm cl}$ of the cloud from the neutron star. The characteristic proton
cooling times and the cloud's radius are also shown. For the cloud mass
supposed in this case one could expect a pulsed emission only during the
first $\leq 1\,h$, since at $t=3\,\rm h$ the gas density in the cloud drops
below $10^{15}\,\rm cm^{-3}$ so that $t_{\rm pp}> P_0$. Note,
however, that calculations in Fig.~\ref{Fig:Arm5} assume an external
pressure free expansion of the cloud with the sound speed. 

In fact the expansion of a fast cloud may be significantly confined due to
the ram pressure of the external medium $p_{\rm ram}= m_{\rm p} n_{\rm gas}
v_{\rm cl}^2 $. For a gas density in the wind of the O-star
declining as $R^{-2}$, the maximum distance
scale ($R_{\rm conf}$) in the binary up to which the ram pressure will be able
to confine the cloud's expansion can be estimated from the condition
$p_{\rm ram} \geq p_{\rm rel}$, where $p_{\rm rel}= w_{\rm rel}/3$ is the
pressure of relativistic particles. For 
$w_{\rm rel} \geq 10^{4}\,\rm erg s^{-1}$ estimated above this results in  
\begin{equation}  
R_{\rm conf}\leq 5\times 10^{12}\, \dot{M}_{-6}^{1/2}   
\left( \frac{v_{\rm cl}}{10^4\,\rm km s^{-1}} \right) \,\rm cm \; , 
\end{equation}  
where we have assumed $v_{\rm w}=10^3\,\rm kms^{-1}$ for the wind speed.
Thus, for a mass loss rate 
$\dot{M} \leq 3\times 10^{-6}\, M_{\odot} \rm yr^{-1}$
the source confinement time
\begin{equation}  
t_{\rm conf}= \frac{R_{\rm conf}}{v_{\rm cl}} \leq  5\times 10^{3}
\dot{M}_{-6}^{1/2}\, \rm s \, ,   
\end{equation}  
does not exceed a couple of hours, {\it independently} of the source speed.
However, a high speed of source is important in order to provide 
$R_{\rm conf} \gg 10^{12}\,\rm cm$ because otherwise the flux of 
pulsed VHE $\gamma$-rays would be effectively suppressed. Eq.(10) then 
suggests that $v_{\rm cl} \sim 10^4\,\rm 
km s^{-1}$ or larger. Note that both Eqs.(10) and (11) limitations relate not
only to the hadronic model but in particular to the leptonic model for
interpretation of pulsed $\gamma$-ray emission considered below in this
section.     

An interesting feature in Fig.~\ref{Fig:Arm5} is a profound absorption of 
$\gamma$-radiation in the thermal UV radiation field of the cloud 
itself, and not only from that of the O-star. It leads to an 
unusually hard spectrum of VHE $\gamma$-rays predicted by this model 
at {\it all} initial stages $t \leq (1-2) \,\rm h$ when the pulsed signal is
produced. In this regard, it is worth remembering that the most
significant indications for a pulsed signal in our 21 February 1999 data are
found in the air shower events corresponding to enhanced $brightness$. 
     
\vspace{3mm} 

Considering now a compact source with a leptonic origin for the pulsed
$\gamma$-radiation, note that in this case we cannot assume that electrons
are accelerated close to the pulsar and then supplied in a relativistic beam 
into the cloud because the life time of HE and VHE electrons in the
radiation field of the companion star (Eq.4), is much shorter than
their travel time to distances $\gg 10^{12}\,\rm cm$ where the source of
VHE $\gamma$-rays should be located, as discussed above.

The principal leptonic model for Cen X-3 then basically corresponds to 
the model developed earlier for the microquasars, i.e. galactic sources with
relativistic jets like GRS 1915+105. For a detailed description of the model
we refer to Atoyan \& Aharonian (\cite{AA99}). Generally, this model assumes
that the inner accretion disc of the compact object in the binary, in our
case the neutron star (or a galactic BH candidate for GRS 1915+105)
sporadically ejects a pair of clouds in two opposite directions to each
other. In the case of stellar-mass black holes the speed of the ejecta may
reach typically $\geq 0.9 c$, whereas in the case of neutron star discs
significantly smaller jet speeds are  assumed (see Mirabel \& Rodriguez
\cite{Mirabel99}). These clouds should then be energized from the central
engine by the relativistic wind (e.g. Poynting flux) propagating in the
jet region and modulated with the pulsar spin period. The latter condition
would not be needed if the pulsed $\gamma$-rays should not be explained. Note
that certain indications that the pairs of ejecta in GRS 1915+105 are
continuously fed with relativistic energy from the central source are found
in the data of radio observations of this microquasar (see Atoyan \&
Aharonian 1999). Relativistic shocks should then be formed at the contact
interface between the ejecta and the wind, providing efficient 
acceleration of electrons. This could result in modulations of the
$\gamma$-ray signal with the pulsar spin period (Doppler shifted), which
would inevitably disappear at times $t\geq t_{\rm conf}$ (Eq.\,11). 

\begin{figure}
\centerline{\epsfig{file=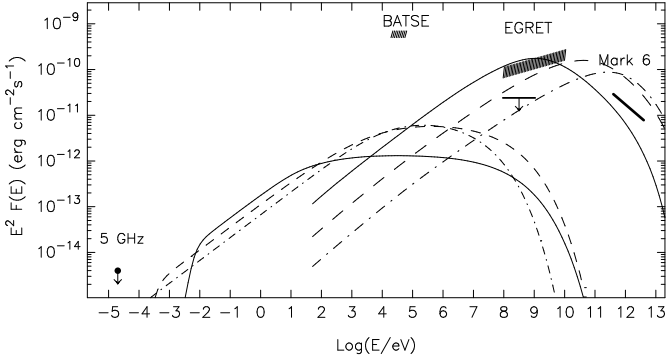, width=8.8cm}}
\caption[]{The spectra of synchrotron (thin lines) and IC (heavy lines)
radiations from a compact fast cloud (`ejecta') propagating in 
the jet(s) of Cen X-3 calculated for 3 different times after ejection:
$t=100\,\rm s$ (solid lines),  $t=1\,\rm h$ (dashed lines),
and $t=1\,\rm d$ (dot-dashed lines). The total injection power in 
shock-accelerated electrons $P_{\rm acc} =10^{37} \,\rm erg s^{-1}$ is 
assumed. The heavy bar at TeV energies corresponds to the mean
integral flux of TeV $\gamma$-rays detected with the Mark-6 telescope in 
1997 (see Chadwick et al. 2000) assuming a power-law differential spectrum 
with $\alpha = 2.6$.}
\label{Fig:Arm6}  
\end{figure}

In Fig.~\ref{Fig:Arm6} we show the broad-band spectra of synchrotron 
and IC radiation predicted by this model at 3 different times after ejection
of the cloud(s) from the accretion disc: $t=100\,\rm s$ (solid
curves), $t=1\,\rm h$ (dashed), and $t= 1\,\rm day$ (dot-dashed). For
calculations in Fig.~\ref{Fig:Arm6} we have formally assumed that the cloud
expands with a mean speed $\simeq 10^{7}\,\rm cm/s$ so that the cloud radius
at $t \sim 1\,\rm h$ would still be  $< cP_0/2$. The magnetic field
behavior in the expanding cloud is approximated as $B(R)\propto R^{-1.5}$,
with a normalisation to $B_0= 50 \,\rm G$ at $R_0 = 10^{10}\,\rm cm$. In such
high magnetic fields, especially at the very initial stages when the cloud is
compact, one has to take into account that synchrotron losses
prevent acceleration of electrons effectively beyond the energy 
$E_{\rm max} \simeq 50/ \sqrt{B/1\,\rm G}\, \rm TeV$. This limitation
partially explains an effective suppression of TeV radiation in
Fig.~\ref{Fig:Arm1} at  $t= 100\,\rm s$. It is more important, however, that
the synchrotron self-absorption at the initial stages effectively removes
from the source all sub-millimetre and radio photons, which become
the principal targets for IC production of VHE radiation in such a compact
source.

An important difference between the hadronic and leptonic compact sources is
that the leptonic model does not need a high gas density in the source, and
the mass load in the cloud may be very low. In that case the Coulomb losses
of relativistic electrons are very low, so the cloud will not be heated to
very high temperatures when the absorption of VHE $\gamma$-rays in the cloud
becomes important. Therefore only the absorption in the UV radiation field of
the O-star may affect the fluxes of VHE $\gamma$-rays which we do not
take into account in Fig.~\ref{Fig:Arm5}). For the spectra at 
$t=100\,\rm s$ this absorption can be rather strong, depending on the orbital 
phase of the pulsar. Note that at that times the spectral modifications are
dominated by the absorption of VHE $\gamma$-rays in the thermal UV
radiation field of the cloud itself. At the times $t= 1\,\rm h$, which is the
basic timescale for production of the pulsed VHE radiation, the source moving
with $v_{\rm cl} \geq 10^{4} \,\rm km s^{-1}$ will be at distances far enough
from the O-star that the absorption of VHE $\gamma$-rays is not dramatic.
Therefore the model predicts rather fast evolution of the VHE radiation
spectra at the stage of pulsations. Note that this prediction is also valid
for the hadronic compact source model (see Fig.~\ref{Fig:Arm5}). 

Unlike the hadronic model, however, the leptonic jet model predicts very
significant fluxes at later stages of the flare evolution, 
$t \geq 1 \,\rm d$, when the pulsations disappear. It is also worth
noting that the shape of the spectra in the HE $\gamma$-ray domain in this
model are much harder, close to the mean spectral flux detected by EGRET in
October 1994 (the hatched region in Fig.~10). Besides that, this model can
provide an answer to the question why the $\gamma$-ray signals seem to be
more persistently detected  (although at different times) by the Mark-6
telescope in the VHE domain than by EGRET at high energies (see dot-dashed
curve in Fig.~\ref{Fig:Arm6}). In this regard note also that the heavy bar in 
Fig.~\ref{Fig:Arm6} is the mean VHE $\gamma$-ray flux detected during 1997,
which is higher by a factor of $\sim 2$ than the mean during 1997-1999.

\vspace{3mm}

Finally, in this section we also have to address the question of the
characteristic frequencies of $\gamma$-ray pulsations that are to be
expected. Because the location of the $\gamma$-ray source cannot coincide
with the pulsar, one should not generally expect that the $\gamma$-ray
pulsations would be coincident with the X-ray pulsations. In the
framework of both models considered in this Section we have to expect a
difference between those pulsations because of the so called `double Doppler
effect', which is the frequency shift $\Delta \nu = \nu - \nu_0$ due to
re-emission/reflection of a pulsed signal from a target moving with a speed
{\bf u } (see Aharonian \& Atoyan \cite{AA91}):
\begin{equation}  
\frac{\Delta \nu}{\nu_0} = \frac{u}{c}\, \frac{\cos({\bf k \hat{ } u}) - 
\cos({\bf p \hat{ } u})} {1-(u/c) \cos({\bf k \hat{ } u}) } \; ,
\end{equation}  
where $u \equiv | \bf u |$ is the cloud's velocity relative to the pulsar;
${\bf k \hat{ } u}$ and ${\bf p \hat{ } u}$ are the angles that the vector
${\bf u}$ makes respectively with the $\gamma$-ray ($\bf k$) and the
relativistic energy outflow ($\bf p$). Then for a compact source moving
at high speed $v_{\rm cl} \sim 10^{4} \,\rm km s^{-1}$ or more favoured by
both models above, an interpretation of the shift of the Rayleigh power peak
at $2.400 \,\rm s$ from the nominal half-period (second harmonic) of the
X-ray pulsar $P_0/2 = 2.410\,\rm s$ would imply that both angles 
${\bf k \hat{ } u}$ and ${\bf p \hat{ } u}$ were rather close, within $\leq
10^\circ$, to $90^\circ$. In principle, for Cen X-3 where the orbital plane
of the pulsar is close to the plane of sky this may still be possible if the
jet is produced almost perpendicular to the orbital plane. However an
interpretation of the $\gamma$-ray pulsations {\it precisely} at the  X-ray
pulsar frequency as reported by Vestrand et al. (\cite{V97}) appears rather 
problematic. If confirmed by future observations, these pulsations
could be reasonably explained only by a model assuming production of HE
radiation very close to the pulsar which should then address again the
problems of the energetics and confinement of such a source.    

\section{Conclusions}

The analysis of the data of observations of Cen X-3 accumulated during 23
days of observations of Cen X-3 with the University of Durham Mark-6 imaging
telescope during 1997-1999 confirms the presence of a $\gamma$-ray signal at
the overall significance level $\geq 4.5\,\sigma$ in agreement with our
previous results (Chadwick et al. \cite{C00}). The behaviour of the signal,
which does not drop but rather increases for images with enhanced brightness
(see Table 1), indicates that the spectra of VHE $\gamma$-rays from Cen X-3
are probably much harder than the spectra of ambient cosmic rays with
$\alpha_{\rm CR} \approx 2.7$. 

No indications for any correlation of the VHE signal with the orbital 
motion of the pulsar, including the pulsar eclipse phase, is found.
Given that a similar result is found by EGRET for HE $\gamma$-rays
detected from Cen X-3, this effect excludes any close vicinity of the pulsar 
as a possible site for $\gamma$-ray production. 

The analysis of the shower arrival times does not reveal any 
statistically significant peak of Rayleigh powers in most of the data, 
except for 1 observation day, the 21 Feb 1999, showing a strong peak with an
estimated probability of occurence of such a peak by chance, after taking
into account the large number of IFF trials, of significantly below 
$10^{-2}$. The analysis using Bayesian statistics results in an even lower
final probability,  $p_{\rm Bay} \simeq 7.5\times 10^{-4}$. 

It is indicative that the peak powers in both statistics are found in the 
data after application of soft image cuts, whereas they completely disappear
when the full (hard) image cuts maximizing significance in the
overall data of observations are applied. This behaviour is just what one
should expect assuming that the modulated signal in the data of 21 Feb 1999
is really connected with VHE $\gamma$-rays. The pulsed signal significantly
increases with the threshold energy of the events, which indicates that the
spectrum of pulsed events is hard as well.      

The position of modulations found in 21 February 1999 data is blueshifted
from the nominal {half period of the X-ray pulsations} by
$\delta \nu /\nu \simeq 3.7 \times 10^{-3}$. Such shifts should be reasonably
expected in any model where the $\gamma$-rays are produced far away from the
pulsar. Moreover, on the basis of the theoretical modelling in this paper,
which suggests high speeds of compact source(s) needed for production of
pulsed VHE $\gamma$-ray emission, we would
generally expect that $\gamma$-ray pulsations could show even  
larger shifts, up to $|\Delta \nu | /\nu_0 \sim 0.3$ in the case of
ejecta moving with the speed $\leq 0.3 c$ such as in SS~433.

The principal models for $\gamma$-ray production in Cen X-3
should be able to produce VHE radiation at distances significantly beyond the
orbit of the pulsar, effectively at $R\geq 3\times 10^{12}\,\rm cm$.
Meanwhile, high energies needed for the $\gamma$-ray fluxes in this binary
can be provided only by the gravitational potential of the neutron star.
Confirmation of the $\gamma$-ray fluxes from Cen X-3 by future observations
would therefore confirm production of powerful jets by the inner  accretion
disc of this X-ray binary. 

Except for the phenomenon of $\gamma$-ray pulsations (which obviously
needs confirmation by forthcoming $\gamma$-ray detectors), an acceptable 
interpretation of the average fluxes currently reported in HE and VHE 
$\gamma$-ray regions can in principle be provided in the framework
of both spatially extended and compact source models, which may also
`co-exist' operating together. The principal difference between these 2 model
approaches is that the extended $\gamma$-ray source implies  
quasi-stationary emission on times scale of at least several days, and
probably even weeks, whereas both hadronic and leptonic compact source models
predict fast evolution of the $\gamma$-ray fluxes on time scales of a few
hours, and with the `disappearance' of the source in a few days. Another
informative feature could be the behaviour of the spectral fluxes at both
high and very high energies.

An important prediction of both compact source models is that pulsed
$\gamma$-ray emission can {\it only} be episodic, with a typical duration
of no more than a few hours.  For $\gamma$-rays of $E \sim 100 \,\rm GeV$
which are to be produced at large distances this is practically a
model-independent requirement. However, $\gamma$-rays with $E\leq 10\,\rm
GeV$ can escape from the binary, even being produced relatively (but not
very!) close to the X-ray pulsar. At these energies, therefore, detection of
pulsed $\gamma$-rays in the eclipse phase of the X-ray pulsar would be
very informative.  

\vspace{1mm}

We can expect that future observations of Cen X-3 and other X-ray
binaries with forthcoming sensitive $\gamma$-ray detectors in the 
$\geq 50 \,\rm MeV$ (GLAST) and $\geq 50\,\rm GeV$ (HESS, CANGAROO-III,
VERITAS) domains will provide large amounts of key information about high
energy processes in these powerful galactic objects.

\begin{acknowledgements}  
We are grateful to the UK Particle Physics and Astronomy Research Council for
support of the project. AMA appreciates the hospitality of the Durham
University during his visit to UK, and of the McGill University where his
contribution to this paper could have been done. The work of AMA has been
supported by the Royal Astronomical Society visitor grant. This paper uses
quick look results provided by the BATSE team.  
\end{acknowledgements}

\end{document}